
%
%
\documentstyle{amsppt}
\magnification=\magstephalf
 \addto\tenpoint{\baselineskip 15pt
  \abovedisplayskip18pt plus4.5pt minus9pt
  \belowdisplayskip\abovedisplayskip
  \abovedisplayshortskip0pt plus4.5pt
  \belowdisplayshortskip10.5pt plus4.5pt minus6pt}\tenpoint
\pagewidth{6.5truein} \pageheight{8.9truein}
\subheadskip\bigskipamount
\belowheadskip\bigskipamount
\aboveheadskip=3\bigskipamount
\catcode`\@=11
\def\output@{\shipout\vbox{%
 \ifrunheads@ \makeheadline \pagebody
       \else \pagebody \fi \makefootline
 }%
 \advancepageno \ifnum\outputpenalty>-\@MM\else\dosupereject\fi}
\outer\def\subhead#1\endsubhead{\par\penaltyandskip@{-100}\subheadskip
  \noindent{\subheadfont@\ignorespaces#1\unskip\endgraf}\removelastskip
  \nobreak\medskip\noindent}
\outer\def\enddocument{\par
  \add@missing\endRefs
  \add@missing\endroster \add@missing\endproclaim
  \add@missing\enddefinition
  \add@missing\enddemo \add@missing\endremark \add@missing\endexample
 \ifmonograph@ 
 \else
 \vfill
 \nobreak
 \thetranslator@
 \count@\z@ \loop\ifnum\count@<\addresscount@\advance\count@\@ne
 \csname address\number\count@\endcsname
 \csname email\number\count@\endcsname
 \repeat
\fi
 \supereject\end}
\catcode`\@=\active
\CenteredTagsOnSplits
\NoBlackBoxes
\def\today{\ifcase\month\or
 January\or February\or March\or April\or May\or June\or
 July\or August\or September\or October\or November\or December\fi
 \space\number\day, \number\year}
\define\({\left(}
\define\){\right)}
\define\Ahat{{\hat A}}

\define\CC{{\Bbb C}}

\define\End{\operatorname{End}}

\define\Hom{\operatorname{Hom}}

\define\RR{{\Bbb R}}

\define\Spin{\operatorname{Spin}}

\define\Tr{\operatorname{Tr}}
\define\ZZ{{\Bbb Z}}
\define\[{\left[}
\define\]{\right]}
\define\ch{\operatorname{ch}}
\define\chiup{\raise.5ex\hbox{$\chi$}}
\define\cir{S^1}
\define\coker{\operatorname{coker}}

\define\exertag #1#2{\removelastskip\bigskip\medskip\eightpoint\noindent%
\hbox{\rm\ignorespaces#2\unskip} #1.\ }

\define\index{\operatorname{index}}

\define\inv{^{-1}}
\define\mstrut{^{\vphantom{1*\prime y}}}
\define\protag#1 #2{#2\ #1}

\define\res#1{\negmedspace\bigm|_{#1}}
\define\temsquare{\raise3.5pt\hbox{\boxed{ }}}

\define\theprotag#1 #2{#2~#1}

\define\zmod#1{\ZZ/#1\ZZ}

\redefine\Re{\operatorname{Re}}

\define\rem{\medskip\noindent{\sl Remark.\/}\ }
\redefine\endrem{\medskip}
\hyphenation{Hb LW iso-metric diff-er-ent}

\NoRunningHeads
\define\property#1#2{{\leftskip30pt\noindent\llap{(#1)\enspace\enspace}#2\par}}
\define\Det{\operatorname{Det}}
\define\Dnor{\dfrac{D\db}{\sqrt{D\db^2}}}
\define\Hb{H_{\bX}}
\define\Ker{\operatorname{Ker}}
\define\LIM{\operatornamewithlimits{LIM}}
\define\Lb{L\db}
\define\Pin{\operatorname{Pin}}
\define\Yg{Y_\gamma}
\define\alim{\operatorname{a-lim}}
\define\bX{\partial X}
\define\cirbdd{\cir_{\text{bounding}}}
\define\cirnonbdd{\cir_{\text{nonbounding}}}
\define\comp#1#2{\left[ #1 \right]_{(#2)}}
\define\curv#1{\Omega^{#1}}
\define\cut{^{\text{cut}}}
\define\dai#1{{\eightpoint #1}}
\define\db{_{\bX}}
\define\etD{e^{-tD_u^2}}
\define\ev{\operatorname{ev}}
\define\gb{\bar{\gamma}}
\define\gtimes{\mathbin{\hat{\otimes}}}
\define\id{\operatorname{id}}
\define\moo{[-1,1]}
\define\path{\Cal{P}Z}
\define\phm{\phi ^-}
\define\php{\phi ^+}
\define\pt{\operatorname{pt}}
\define\ra{\longrightarrow}
\define\sign{\operatorname{sign}}
\define\sind{(-1)^{\index D_Y}}
\define\spec{\operatorname{spec}}
\define\tpi{2\pi i}
\define\zo{[0,1]}

\refstyle{A}
\widestnumber\key{SSSSSSS}   

 \pretitle{$$\boxed{\boxed{\text{REVISED VERSION}}}$$\par\vskip 3pc}

        \topmatter
 \title\nofrills $\eta $-Invariants and Determinant Lines \endtitle
 \author Xianzhe Dai\\Daniel S. Freed  \endauthor
 \thanks The first author is supported by NSF grant DMS-9204267 and Alfred P.
Sloan Fellowship. He would also like to thank MSRI for its support and
hospitality while some of this work was completed.  The second author is
supported by NSF grant
DMS-8805684, a Presidential Young Investigators award DMS-9057144, and by the
O'Donnell Foundation.  He would also like to thank the Geometry Center at the
University of Minnesota for its hospitality while some of this work was
completed.\newline \indent To appear in the {\sl Journal of Mathematical
Physics\/} in the special issue on {\it Topology and Physics\/}.\endthanks
 \affil Department of Mathematics, University of Southern California \\
 Department of Mathematics, University of Texas at Austin\endaffil
 \address Department of Mathematics, University of Southern California, Los
Angeles, CA 90089\endaddress
 \email xdai\@math.usc.edu \endemail
 \address Department of Mathematics, University of Texas, Austin, TX
78712\endaddress
 \email dafr\@math.utexas.edu \endemail
 \date May 26, 1994\enddate
        \endtopmatter

\document

The $\eta $-invariant was introduced by Atiyah, Patodi, and Singer~\cite{APS}
in a series of papers treating index theory on {\it even\/} dimensional
manifolds with boundary.  It first appears there as a boundary correction in
the usual local index formula.  Suppose $X$~is a closed {\it odd\/}
dimensional spin manifold (which in their index theorem is the boundary of an
even dimensional spin manifold).  The Dirac operator\footnote{For simplicity
we only consider the basic Dirac operator, though as usual in geometric index
theory all of our results hold for twisted Dirac operators, i.e., for
operators of ``Dirac-type''.}~$D_X$ is self-adjoint and has discrete real
spectrum.  Define
  $$ \eta _X(s) = \sum\limits_{\lambda \not= 0}\frac{\sign\lambda }{|\lambda
     |^s},\qquad \Re(s)>>0, $$
where the sum ranges over the nonzero spectrum of~$D_X$.  Then $\eta
_X(s)$~is analytic in~$s$ and has a meromorphic continuation to~$s\in \CC$.
It is regular at~$s=0$, and its value there is the $\eta $-invariant.  More
precisely, what appears in the Atiyah-Patodi-Singer index formula is the {\it
$\xi $-invariant\/}
  $$ \xi _X=\frac{\eta _X(0)+\dim\ker D_X}{2}. $$
Under a smooth variation of parameters (for example, the metric on~$X$) the
$\xi $-invariant jumps by integers, whereas $\xi \pmod1$~is smooth.  In this
paper we are interested in the latter, so consider the {\it exponentiated\/}
$\xi $-invariant
  $$ \tau _X=e^{\tpi\xi _X} $$
instead.  In fact, our interest is in manifolds with boundary and we use
``global'' self-adjoint elliptic boundary conditions for the Dirac operator
which are the odd dimensional analog of the Atiyah-Patodi-Singer boundary
conditions~\cite{APS}.  To formulate these boundary conditions we need to
choose a ``trivialization'' of the graded kernel of the Dirac operator
on~$\bX$.\footnote{Other authors describe this choice as a Lagrangian
subspace of the kernel, or as an involution on the kernel.  All of these
descriptions are equivalent.} The exponentiated $\xi $-invariant depends on
this trivialization (\theprotag{1.4} {Theorem}) in a simple way.

Our first observation is that this dependence means that the exponentiated
$\xi $-invariant naturally lives in the {\it inverse\/}\footnote{An
unfortunate choice of sign in the whole index theory----perhaps dating back
to Fredholm---explains why it is the {\it inverse\/} determinant line which
occurs here.  An operator $D\:H^+\to H^-$ is an element of~$H^-\otimes
(H^+)^*$, so the codomain appears with a $+$~sign and the domain with a
$-$~sign.  It would be better, then, to define the index of~$D$ as
$\dim\coker D - \dim\ker D$.  To make the index theorem for manifolds with
boundary come out, the $\xi $-invariant would also be defined with the
opposite sign from the usual one, as would the $\Ahat$-genus.  On the other
hand, the determinant line~\thetag{2.7} is defined with the ``proper'' sign.
Regardless of what is proper, this discrepancy explains some of the funny
signs which crop up in index theory.} determinant line of the Dirac operator
on the boundary (\theprotag{2.15} {Proposition}).  In fact, it has unit norm
in the Quillen metric.  For a family of Dirac operators this invariant is
then a section of the inverse determinant line bundle over the parameter
space.  In \theprotag{1.9} {Theorem} we generalize the usual formula for the
variation of the $\xi $-invariant to a formula for the {\it covariant
derivative\/} of this section.  Here we use the natural connection on the
(inverse) determinant line bundle defined by Bismut and Freed~\cite{BF1}.
The proof of \theprotag{1.9} {Theorem} occupies~\S{3}.  Our other main result
is a {\it gluing formula\/} for the exponentiated $\xi $-invariant, which we
state in \theprotag{2.20} {Theorem} and prove in~\S{4}.  To get the signs
right in that theorem we view the determinant line as a {\it graded\/} vector
space, as explained in~\S{2}.  In~\S{5} we give a new proof of the holonomy
formula for the natural connection on the determinant line bundle~\cite{BF2},
{}~\cite{C2}.  This formula was originally conjectured by Witten~\cite{W} in
connection with global anomalies.  It expresses the holonomy, or global
anomaly, as the adiabatic limit of an exponentiated $\xi $-invariant.
In~\S{6} we explain how our results lead to a conjecture about the
geometrical index of families of Dirac operators on odd dimensional manifolds
with boundary.\footnote{We understand that ongoing work of Melrose and Piazza
is expected to prove this conjecture.}

Our results build on previous work treating $\eta $-invariants on manifolds
with boundary.  Many different kinds of boundary conditions appear in these
works.  Cheeger~\cite{C1,\S6} introduced the $\eta $-invariant (for the
signature operator) on manifolds with conical singularities, and he notes
that this corresponds to global boundary conditions on a manifold with
boundary when one attaches a cone to the boundary.  Further, his ``ideal
boundary conditions'' correspond to the trivialization of the graded kernel
on the boundary.  In later work~\cite{C2} he proves a variational formula for
the $\eta $-invariant on a manifold with conical singularities.  Gilkey and
Smith~\cite{GS} discuss the $\eta $-invariant for {\it local\/} boundary
conditions, which were used in the original proof of the Atiyah-Singer index
theorem to show that the index is a bordism invariant~\cite{P}.
Singer~\cite{Si} proved a formula relating the difference of $\eta
$-invariants for two specific local boundary conditions with the determinant
of the laplacian on the boundary.  Mazzeo and Melrose~\cite{MM} assume that
the boundary Dirac operator is invertible and then define an $\eta
$-invariant using Melrose's ``$b$-calculus''.  With this assumption they
prove a gluing law.  Dai~\cite{D1} proved a formula relating this ``$b$-eta
invariant'' to the $\eta $-invariant defined with local boundary conditions.
Another approach is to attach a half-cylinder to the boundary and use
$L^2$~spinor fields.  This was considered in special cases by the della
Pietras~\cite{dlP1}, ~\cite{dlP2} and more generally by
Klimek/Wojciechowski~\cite{KW} and M\"uller~\cite{M\"u}.  M\"uller proves
that this $\eta $-invariant is equal to the $\eta $-invariant for the global
boundary conditions with a certain trivialization of the kernel picked out by
the kernel of the Dirac operator on $L^2$~spinor fields.  It is also easy to
see that it agrees with the $b$-eta invariant if the metric near the boundary
is asymptotically cylindrical.  The self-adjoint global boundary conditions,
and certain generalizations, were first studied by Douglas and
Wojciechowski~\cite{DW}.  M\"uller~\cite{M\"u} gives a systematic treatment
of the analytic aspects of these self-adjoint boundary conditions.  Lesch and
Wojciechowski~\cite{LW} determine the dependence of the exponentiated $\xi
$-invariant on the boundary trivialization (\theprotag{1.4} {Theorem}).
M\"uller~\cite{M\"u} derives this result as well; his argument rests on a
variational formula.  Bunke~\cite{B1} proves a gluing formula for the
(unexponentiated) $\eta $-invariant in case a closed manifold is split into
two pieces.  Recent preprints of Wojciechowski~\cite{Wo1}, ~\cite{Wo2} also
prove gluing formulas for the $\eta $-invariant modulo one.

Our contribution here begins with our geometric formulation of the
exponentiated $\xi $-invariant as taking values in the inverse determinant
line.  For example, this leads to a geometric variational
formula~\thetag{1.10} which is crucial in all of our subsequent work.  In
particular, the variational formula relates the exponentiated $\xi
$-invariant to the natural connection on the determinant line bundle.  The
gluing law we prove (\theprotag{2.20} {Theorem}) is more general than that
obtained by cutting a closed manifold into two pieces.  This is necessary for
example in~\S{5}, where we glue together cylinders.  Thus we must consider
gluing along manifolds where the index of the Dirac operator may be nonzero.
The most natural formulation of the result is in terms of {\it graded\/}
determinant lines.  This notion is discussed in Knudsen and Mumford~\cite{KM}
who credit the idea to Grothendieck.  It also appears in later work of
Deligne~\cite{Del} as clearly the best way to avoid a {\it cauchemar de
signes\/}!  Our proof of the gluing law in~\S{4} is simpler than previous
proofs.  We begin with the same patching of spinor fields as in
Bunke~\cite{B1}.  Then we note a symmetry which allows us to conclude easily
with the variation formula.  It is tempting to speculate that this approach
to gluing may be useful in other linear problems and in nonlinear problems as
well.

Our proof of the holonomy theorem---also known as the {\it global anomaly
formula\/}---is considerably simpler than previous proofs, partly due to our
simple proof of the gluing law.  We rely heavily on geometric ideas.  Thus we
avoid any consideration of large time behavior of heat kernels, and we also
avoid using non-pseudodifferential operators~\cite{BF2}.  Cheeger's argument
in~\cite{C2,\S9}, which proves the adiabatic limit formula for the signature
operator in the invertible case, bears much resemblance to our proof here.
He works on a manifold with conical singularities and applies his variational
formula and his ``singular continuity method''; the latter is analogous to
our use of gluing.  The idea of considering parallel transport also appears
in papers of the della Pietras~\cite{dlP1}, ~\cite{dlP2}, but they failed to
consider gluing.  Our proof proceeds as follows: We use gluing to show that
the adiabatic limit of exponentiated $\xi $-invariants on cylinders defines
the parallel transport of a connection on the determinant line bundle.  Then
we apply our geometric variational formula to prove that it agrees with the
natural connection.  In a sense we use the gluing law to break up the
holonomy---a global problem on the circle---into a composition of parallel
transports---local problems on small intervals.

The idea of computing global invariants on closed manifolds from local
invariants on manifolds with boundary using gluing laws is informed by recent
work in {\it quantum field theory\/}, particularly {\it topological\/}
quantum field theory.  The gluing is a characteristic property of the path
integral, and it follows formally from a similar property of the classical
action.  These gluing laws are fundamental for computing quantum Chern-Simons
invariants, Donaldson polynomials, and other topological and geometric
invariants.  Older invariants in topology and geometry also obey gluing
laws~\cite{F3}, ~\cite{F4} and our work here fits the $\eta $-invariant into
this story.  The theory of the {\it classical\/} Chern-Simons
invariant~\cite{F2} is very similar, and of course the original papers of
Atiyah, Patodi, and Singer~\cite{APS} discuss the relationship of $\eta
$-invariants (and so exponentiated $\xi $-invariants) to Chern-Simons
invariants for closed manifolds.  We also remark that certain ratios of
exponentiated $\xi $-invariants are topological invariants which live in
$K\inv $-theory with $\RR/\ZZ$~coefficients \cite{APS}.  Our work gives a
factorization of these topological invariants as well.  It is tempting to say
that the exponentiated $\xi $-invariant is {\it local\/} and so can serve as
an action for a field theory, just as the Chern-Simons invariant can.  (For
example, see the recent preprint~\cite{B2}.)  One crucial difference is that
the Chern-Simons invariant is multiplicative in coverings, whereas the
exponentiated $\xi $-invariant is {\it not\/}.  In any case, the gluing law
does exhibit some local properties of the $\eta $-invariant.

The suggestion that the $\eta $-invariant of a (3-)manifold with boundary
lives in the determinant line of the boundary was made in a manuscript of
Graeme Segal~\cite{S}.  We thank Segal for sharing his ideas with us.  We
also benefited from conversations with Ulrich Bunke and John Lott.

\newpage
\head
\S{1} The exponentiated $\xi $-invariant
\endhead
\comment
lasteqno 1@ 10
\endcomment

Suppose $X$~is a compact odd dimensional spin\footnote{We understand a spin
manifold to have a definite metric, orientation, and spin structure.  Our
work extends to $\text{spin} ^c$ manifolds and to Dirac operators twisted by
a vector bundle with connection, but for simplicity we omit these
refinements.} manifold with nonempty boundary.  Assume that $X$~has a metric
with an explicit product structure near~$\bX$.  Thus in a neighborhood of the
boundary there is a given isometry with~$(-1,0]\times \bX$.  Let $H_X$~denote
the Hilbert space of $L^2$~spinor fields on~$X$ and $D_X\:H_X\to H_X$ the
formally self-adjoint Dirac operator.  We use similar notation for the
induced Dirac operator on the boundary.

Our first job is to specify self-adjoint elliptic boundary conditions.  Our
discussion here is somewhat formal.  We leave the detailed analysis to the
appendix.  Let $J\:\Hb\to \Hb$ be Clifford multiplication by the {\it
outward\/} unit normal vector field to the boundary.  Then $J$~is
skew-adjoint, $J^2=-1$, and $D_{\bX}J=-JD_{\bX}$.  The $\mp i$-eigenspaces
of~$J$ induce the usual splitting $H\mstrut \db = \Hb^+ \oplus \Hb^-$.  Now
integration by parts yields the formula
  $$ (D_X\psi ,\varphi )_X - (\psi ,D_X\varphi )_X = (J\psi ,\varphi
     )_{\bX},\qquad \psi ,\varphi \in H_X.  $$
Thus if our boundary condition is described by $\psi \res{\bX}\in W\subset
\Hb$, then the corresponding Dirac operator is self-adjoint if $JW =
W^{\perp}$, at least formally.  We also need {\it elliptic\/} boundary
conditions, so we choose~$W$ ``close'' to the subspace which describes the
Atiyah-Patodi-Singer nonlocal boundary \break conditions~\cite{APS}.

Our precise choice is this.  The nonnegative self-adjoint
operator~$D^2_{\bX}$ induces decompositions
  $$ H\db^\pm = K\db^\pm \oplus \bigoplus_{\lambda >0}E\db^\pm (\lambda ),
      $$
where $K\db^+\oplus K\db^-$ is the kernel of~$D\db$ and $E\db^+(\lambda
)\oplus E\db^-(\lambda )$ is the eigenspace with eigenvalue~$\lambda $.  The
sum is over the spectrum~$\spec(D\db^2)$.  Note that
  $$ D\mstrut \db\:E\db^+(\lambda )\longrightarrow E^-\db(\lambda )
      $$
is an isomorphism, though it is not unitary---it is $\sqrt\lambda $~times a
unitary map.  Also, by the cobordism
invariance of the index~\cite{P} we have $\index D\db =0$ and so $\dim K\db^+
= \dim K\db^-$.  Now for any positive $a\notin \spec(D\db^2)$ let
  $$ \aligned
     K\db^\pm(a) &= K\db^{\pm }\oplus \bigoplus_{0< \lambda <a}
E\db^\pm(\lambda
     )\\
     H\db^\pm(a) &= \bigoplus_{\lambda >a} E\db^\pm(\lambda ).\endaligned
     \tag{1.1} $$
By ellipticity $K\db^{\pm}(a)$~is finite dimensional.
A choice of boundary condition~$W_{\langle a,T  \rangle}$ is determined by
the number~$a$ and by a choice of {\it isometry\/}
  $$ T\:K\db^+(a)\longrightarrow K^-\db(a).  $$
Let $\Dnor$~denote the operator which restricts to~$D_{\bX}/\sqrt{\lambda }$
on~$E\db^+(\lambda )$; it is defined on~$H\db^+ \ominus K\db^+$.  We denote
its restriction to~$H^+_{\bX}(a)$ by~$\dfrac{D_{\bX}(a)}{\sqrt{D_{\bX}(a)^2}}$.
 A spinor field~$\phi
^+\in H\db^+$ decomposes according to $H\db^+ = K\db^+(a)\oplus H\db^+(a)$.
Then
  $$ W_{\langle a,T \rangle}= \left\{ \langle \phi ^+,\phi ^- \rangle \in
     H\db : \phi ^- + \bigl(T\oplus \dfrac{D\db(a)}{\sqrt{D\db(a)^2}}
     \bigr)\phi ^+ =0\right\}.  \tag{1.2} $$
This is a generalization of the boundary condition studied by previous
authors\footnote{Other authors describe the isometry~$T$ by its graph, which
is a {\it lagrangian\/} subspace of the kernel.} (\cite{DW}, \cite{LW},
\cite{B1}, \cite{M\"u}), who choose~$a$ less than the first eigenvalue
of~$D\db^2$.  We need this generalization to treat families.

Now for any choice~$\langle a,T \rangle$ of boundary conditions the Dirac
operator~$D_X(a,T)$ is self-adjoint elliptic and has a well-defined $\eta
$-invariant $\eta _X(a,T)$.  (See Appendix.)  We use the more refined $\xi
$-invariant
  $$ \xi _X(a,T) = \frac{\eta _X(a,T) + \dim\ker D_X(a,T)}{2}  $$
and set
  $$ \tau _X(a,T) = e^{\tpi \xi _X(a,T)}.  $$
Our first result is a generalization of~\cite{LW}, ~\cite{B1,Corollary 9.3},
and~\cite{M\"u,Theorem 2.21}.  It computes the dependence of~$\tau _X(a,T)$
on~$\langle a,T \rangle$.  To state it note that if $0<a<b$ with $a,b\notin
\spec(D\db^2)$, and $T\:K\db^+(a)\to K\db^-(a)$ is an isometry, then $T\oplus
\dfrac{D\db(a,b)}{\sqrt{D\db(a,b)^2}} \:K\db^+(b)\to K\db^-(b)$ is also a
unitary isomorphism.  Here $D\db(a,b)$~denotes the restriction of~$D\db$ to
  $$ H\db^\pm(a,b) = \bigoplus_{a<\lambda <b} E\db^\pm(\lambda ). \tag{1.3}
     $$

        \proclaim{\protag{1.4} {Theorem}}
 Suppose $0<a<b$ with $a,b\notin \spec(D\db^2)$ and $T,T\mstrut _1,T\mstrut
_2\:K\db^+(a)\to K\db^-(a)$ are isometries.  Then
  $$ \align
      \tau _X(a,T\mstrut _2) &= \det(T_1\inv T\mstrut _2) \,\tau
     _X(a,T\mstrut _1), \tag{1.5} \\
      \tau _X(b,T\oplus \dfrac{D\db(a,b)}{\sqrt{D\db(a,b)^2}}) &= \tau
     _X(a,T). \tag{1.6}\endalign $$
        \endproclaim

 \flushpar
 Equation~\thetag{1.6}~is trivial since $W_{\langle b,T\oplus D\mstrut
\db(a,b)/\sqrt{D\db(a,b)^2} \rangle} = W_{\langle a,T \rangle}$.  We defer
the proof of~\thetag{1.5} to~\S{4} ({\theprotag{4.22} {Corollary}}).

We can interpret~\thetag{1.5} and~\thetag{1.6} as instructions for
constructing a hermitian line~$L\db$ and an element~$\tau _X\in L\db$.
Namely, let $\Cal{C}\db = \{\langle a,T  \rangle\}$ be the set of possible
boundary conditions and then define the complex line
  $$ L_{\bX} = \{\tau \:\Cal{C}\db\to\CC : \tau \text{ satisfies~\thetag{1.5}
     and \thetag{1.6}} \}. \tag{1.7} $$
Since $|\det(T_1\inv T\mstrut _2)|=1$ in~\thetag{1.5} we see that the
expression
  $$ (\tau _1,\tau _2) = \tau _1(a,T)\overline{\tau_2(a,T)}  $$
is independent of~$\langle a,T  \rangle$ and so defines a hermitian metric
on~$L\db$.  By construction $\tau _X\in L\db$ is an element of unit norm.

We use a patching construction to extend to families (cf.~\cite{F1}).  Let
$\pi \:X\to Z$ be a fiber bundle whose typical fiber is a compact odd
dimensional manifold with boundary, and let $\partial \pi \:\bX\to Z$ be the
fiber bundle of the boundaries.  A Riemannian structure on $X\to Z$ is a
metric on the relative tangent bundle~$T(X/Z)$ together with a field of
horizontal planes on~$X$, which we specify as the kernel of a projection
$P\:TX\to T(X/Z)$.  Suppose also that $T(X/Z)$~is endowed with an orientation
and spin structure.  For simplicity we term~$\pi $ a `spin map'.  For our
purposes we also assume that the metrics are products near the boundaries.
Now for each~$a>0$ define
  $$ U_a=\{z\in Z: a\notin \spec(D_{\bX_z}^2)\}.  $$
On this open set $K_{\bX_z}^\pm(a)$ are smooth vector bundles of equal rank.
Choose a cover
  $$ U_a = \bigcup_{i} U_{a,i} \tag{1.8} $$
so that these bundles are isomorphic over each~$U_{a,i}$.  Then choose a
smooth family of isomorphisms $T_z(a,i)\:K_{\bX_z}^+(a) \to K_{\bX_z}^-(a)$
and compute $\tau _{X_z}\bigl(a,T_z(a,i)\bigr)$, which is a smooth function
of~$z$.  The collection of these functions for various choices of~$a$, $i$,
and~$T_z(a,i)$ satisfy~\thetag{1.5} and~\thetag{1.6}.
Definition~\thetag{1.7} extends to this situation---now everything depends
smoothly on~$z$---to define a hermitian line bundle $L_{\bX/Z}\to Z$.  The
functions~$\tau _{X_z}\bigl(a,T_z(a,i)\bigr)$~patch together to form a smooth
section~$\tau _{X/Z}$ of~$L_{\bX/Z}$.

In~\S{2} we identify~$L_{\bX/Z}$ as the inverse determinant line bundle of
the family of Dirac operators on~$\bX\to Z$ with its Quillen metric.  This
line bundle carries a natural unitary connection\footnote{In~\S{5} we define
a connection~$\nabla '$ directly on~$L_{\bX/Z}$ using the invariant~$\tau
_X$.  We prove that it agrees with~$\nabla $ under the isomorphism with the
inverse determinant line bundle.}~$\nabla $, constructed in~~\cite{BF1}.  The
following theorem computes the covariant derivative of~$\tau _{X/Z}$ with
respect to this connection; it generalizes the standard formula on closed
manifolds (e.g.~~\cite{BF2,Theorem~2.10}).

        \proclaim{\protag{1.9} {Theorem}}
 Let $\pi \:X\to Y$ be a spin map whose typical fiber is an odd dimensional
manifold with boundary.  Let $\Omega ^{X/Z}$~denote the curvature of the
relative tangent bundle and $\Ahat(\Omega ^{X/Z})$~its $\Ahat$-polynomial.
Then the covariant derivative of the exponentiated $\xi $-invariant
is\footnote{In~\thetag{1.10} we use the standard sign convention
(e.g.~\cite{BT}) for integration over the fiber.  For example, if $\alpha
$~is a form on~$Z$ and $\beta $~an $n$-form on an oriented manifold~$X^n$,
then
  $$ \int_{(Z\times X)/Z}\alpha \wedge \beta = \left( \int_{X}\beta
     \right)\,\alpha .  $$
}
  $$ \nabla \tau _{X/Z} = \tpi \int_{X/Z} \comp{\Ahat(\Omega ^{X/Z})}{1}\cdot
     \tau _{X/Z}.  \tag{1.10} $$
        \endproclaim

\flushpar
 We defer the proof to~\S{3}.

\newpage
\head
\S{2} Graded Determinant Lines
\endhead
\comment
lasteqno 2@ 25
\endcomment

Our first goal in this section is to identify the hermitian
line~$L\db$~\thetag{1.7} with the {\it inverse\/}\footnote{The
inverse~$L\inv$ of a one dimensional vector space~$L$ is its dual~$L^*$.}
determinant line~$\Det\inv \db$ of the Dirac operator~$D\db$.  The hermitian
structure on~$\Det\db$ is due to Quillen~\cite{Q}.  We then state various
properties of~$\tau _X$ and~$L\db$, the most important of which is the gluing
law (\theprotag{2.20} {Theorem}).  Here we encounter inverse determinant
lines for operators of nonzero index.  Then the gluing law involves some
signs which are best understood in terms of the {\it grading\/} on the
determinant line given by the index~\cite{KM}.  Hence we begin this section
with an exposition of graded vector spaces.

A {\it graded vector space\/} $V=V^+\oplus V^-$ is simply a direct sum of two
vector spaces, which in this paper we always take to be complex.  We
call~$V^+$ (resp.~$V^-$) the even (resp.~odd) part of~$V$, and write $|v|=0$
(resp.~$|v|=1$) for~$v\in V^+$ (resp.~$v\in V^-$).  For graded vector
spaces~$V,W$ we write $V\gtimes W$ for the graded vector space whose
underlying vector spaces is~$V\otimes W$ and with $|v\otimes w| \equiv |v| +
|w|\pmod2$ for homogeneous elements~$v\in V$, $w\in W$.  We use the
$\gtimes$~notation to keep track of signs in the isomorphism
  $$ \aligned
      V\gtimes W &\longrightarrow W\gtimes V\\
      v\otimes w&\longmapsto (-1)^{|v|\,|w|}\,w\otimes v,\qquad v\in V,\quad
     w\in W.  \endaligned \tag{2.1} $$
Here, as in subsequent expressions, we use homogeneous elements and extend by
linearity.  The dual space $V^*=(V^+)^*\oplus (V^-)^*$ of a graded vector
space is also graded, and we use the natural pairing
  $$ \aligned V^*\gtimes V &\longrightarrow \CC\\ \check v\otimes v
     &\longmapsto \check v(v),\qquad v\in V,\quad \check v\in V^*.
     \endaligned \tag{2.2} $$
The order of the factors in~\thetag{2.2} is important!  With this choice
there is no sign in~\thetag{2.2}, nor is there any in the isomorphisms
  $$ \aligned
     V^*\gtimes W^* &\longrightarrow (W\gtimes V)^*\\
     \check v\otimes \check w &\longmapsto \bigl(\ell \:w\otimes v \mapsto
     \check v(v)\,\check w(w) \bigr)     \endaligned \tag{2.3} $$
and
  $$ \aligned
     W\gtimes V^* &\longrightarrow \Hom(V,W)\\
     w\otimes \check v &\longmapsto \bigl(T\:v\mapsto \check v(v)\,w \bigr).
     \endaligned \tag{2.4} $$
Notice that the natural isomorphism
  $$ \aligned
     V&\longrightarrow V^{**}\\
     v &\longrightarrow \bigl(\ell \:\check v\mapsto (-1)^{|v|\,|\check
     v|}\,\check v (v) \bigr)     \endaligned \tag{2.5} $$
picks up a sign in the graded context.  The sequence of homomorphisms
  $$ \Tr_s\:\End(V) @>\thetag{2.4}>> V\gtimes V^* @>\thetag{2.1}>> V^*\gtimes
     V @>{\dsize\Tr} >> \CC \tag{2.6} $$
is the {\it supertrace\/}: For $T=\left(\smallmatrix A&B\\C&D
\endsmallmatrix\right)\in \End(V^+\oplus V^-)$ we have $\Tr_sT=\Tr A - \Tr
D$.

The {\it determinant line\/}~$\Det V$ of an {\it ungraded\/} vector space~$V$
is the one dimensional vector space of totally antisymmetric tensors $\omega
=v_1\wedge \dots \wedge v_n$.  We view~$\Det V$ as a {\it graded\/} vector
space whose degree is~$\dim V\pmod 2$.  If $V=V^+\oplus V^-$ is graded, then
define
  $$ \Det V= (\Det V^-)\gtimes (\Det V^+)\inv .  \tag{2.7} $$
This is again a graded line, the grading given by
  $$ |\Det V| \equiv \dim V \equiv \dim V^+ - \dim V^-\pmod2.  $$
Using~\thetag{2.4} we see that if $\dim V^+ = \dim V^-$ then the top exterior
power of a homomorphism $T\:V^+\to V^-$ determines an element
  $$ \Det T\in \Det V. \tag{2.8} $$
If $V^+=V^-$ then $T$~has a numerical determinant~$\det T\in \CC$, and this
is related to~\thetag{2.8} via the supertrace~\thetag{2.6}:
  $$ \Tr_s(\Det T) = (-1)^{\dim V^+}\,\det T. \tag{2.9} $$
Let $-V$ denote~$V$ with the opposite grading: $(-V)^\pm = V^\mp$.  Note the
sign in the isomorphism
  $$ \aligned
      \Det(-V)&\longrightarrow \Det(V)\inv \\
      \omega ^+\otimes \check\omega ^- &\longmapsto \bigl(\ell \:\omega
     ^-\otimes \check\omega ^+ \mapsto (-1)^{\dim V^+}\,\check\omega ^+(\omega
     ^+)\;\check\omega ^-(\omega ^-) \bigr),
      \endaligned \tag{2.10} $$
where $\omega ^\pm\in \Det(V^\pm)$ and $\check\omega ^\pm\in \Det(V^\pm)\inv$.
Similarly, if $W$~is another graded vector space, then there is a sign in the
isomorphism
  $$ \aligned
     \Det(V\oplus W) &\longrightarrow \Det V\gtimes \Det W\\
     \omega ^-\otimes \eta ^-\otimes \check\eta ^+\otimes \check\omega ^+
     &\longmapsto (-1)^{\dim V^+\dim W} \,\omega ^-\otimes \check\omega
     ^+\otimes \eta ^-\otimes \check\eta ^+,\endaligned \tag{2.11} $$
where $\omega ^\pm\in \Det(V^\pm)$ and~$\eta ^\pm\in \Det(W^\pm)$.

As a matter of notation, if $\omega \in L$ is a nonzero element of a graded
line~$L$, then we denote by~$\omega \inv \in L\inv $ the unique element so
that $\omega \inv (\omega )=1$ under the pairing~\thetag{2.2}.

Suppose $V,W$~are graded vector spaces with $\dim V^+=\dim V^-$ and $\dim W^+
= \dim W^-$.  Note in particular that $\dim W$~and $\dim V$~are even.  Then
for $T\:V^+\to V^-$ and $S\:W^+\to W^-$ we have
  $$ \gather
      \Det(T\inv ) = (-1)^{\dim V^+}(\Det T)\inv  \\
      \Det(T\oplus S) = \Det T\otimes \Det S. \endgather $$
The equalities here stand for the isomorphisms~\thetag{2.10}
and~\thetag{2.11}.

Next, we review the construction of the determinant line of a Dirac operator
(see~\cite{F1} for details), but now as a {\it graded\/} line.  Let $Y$~be a
closed even dimensional spin manifold.  The spinor fields $H\mstrut _Y =
H^+_Y \oplus H^-_Y$ on~$Y$ are graded, and the Dirac operator $D\mstrut
_Y\:H^+_Y\to H^-_Y$ anticommutes with the grading.  We use the notations
$K_Y(a)$, $H_Y(a)$, and~$H_Y(a,b)$ from~\thetag{1.1} and~\thetag{1.3}, where
$a<b$~are positive numbers not in~$\spec(D_Y^2)$.  Now $D\mstrut
_Y(a,b)=D\mstrut _Y\:H^+_Y(a,b)\to H_Y^-(a,b)$ is an isomorphism, so
  $$ \Det D_Y(a,b)\in \Det H_Y(a,b)  $$
is a nonzero element.  Define an isomorphism
  $$ \aligned
      \theta _Y(a,b)\:\Det K_Y(a) &\longrightarrow \Det K_Y(a)\gtimes \Det
     H_Y(a,b)\cong \Det K_Y(b)\\
      \omega (a)&\longmapsto \omega (a)\otimes \Det D_Y(a,b).  \endaligned
     \tag{2.12} $$
Then the determinant line is defined to be a set of compatible
elements~$\omega (a)\in \Det K_Y(a)$:
  $$ \Det_Y = \bigl\{ \omega =\{\,\omega (a)\in \Det
     K_Y(a)\,\}_{a\notin\spec(D_Y^2)} : \omega (b) = \theta _Y(a,b)\omega (a)
     \bigr\}.  $$
Note that
  $$ |\Det _Y| \equiv \index D_Y\pmod2.  $$
Now the lines $\Det K_Y(a)$ and~$\Det H_Y(a,b)$ inherit hermitian metrics
from the $L^2$~metric on~$H_Y$, and we compute
  $$ |\theta (a,b)\,\omega (a)|^2_{K_Y(b)} = \left(\prod\limits_{a<\lambda
     <b}\lambda \right) |\omega (a)|^2_{K_Y(a)},\qquad \omega (a)\in \Det
     K_Y(a).   $$
Hence the expression
  $$ |\omega |^2_{\Det_Y} = \left( \prod\limits_{\lambda >a}\lambda
     \right)|\omega (a)|^2_{\Det K_Y(a)}  $$
is independent of~$a$, where the product is defined using a $\zeta
$-function.  Equation~\thetag{2.8} defines the {\it Quillen metric\/}
on~$\Det_Y$.

A careful computation shows that \thetag{2.10}~and \thetag{2.11}~are
compatible with the ``patching'' isomorphism~$\theta (a,b)$ in~\thetag{2.12},
so they determine isometries
  $$ \gather
      \Det\mstrut _{-Y}\cong \Det\inv_Y \tag{2.13}\\
      \Det_{Y_1\sqcup Y_2}\cong \Det_{Y_1}\gtimes \Det_{Y_2}.  \tag{2.14}
       \endgather $$
Here $Y,Y_1,Y_2$~are closed spin manifolds, `$-Y$'~denotes the spin
manifold~$Y$ with the opposite orientation,\footnote{Let $\Spin(Y)\to Y$
denote the principal $\Spin_n$~bundle which defines the spin structure
of~$Y$; it is a double cover of the bundle of oriented orthonormal frames.
Then the spin structure on~$-Y$ is defined by the complement of~$\Spin(Y)$ in
the $\Pin_n$~bundle of frames $\Spin(Y)\times _{\Spin_n}\Pin_n\to Y$.} and
`$Y_1\sqcup Y_2$'~denotes the disjoint union of~$Y_1$ and~$Y_2$.

The patching isomorphism used to patch the inverse determinant line (which
appears in~\thetag{2.13}, for example) is
  $$ \aligned
      \bigl(\theta _Y(a,b)^* \bigr)\inv \:\bigl(\Det K_Y(a)\bigr)\inv
     &\longrightarrow \bigl(\Det H_Y(a,b)\bigr)\inv \gtimes \bigl(\Det
     K_Y(a)\bigr)\inv \cong \bigl(\Det K_Y(b)\bigr)\inv \\
      \eta (a)&\longmapsto \bigl(\Det D_Y(a,b)\bigr)\inv \otimes \eta (a).
     \endaligned  $$

With this understood we can identify the hermitian line determined by the
exponentiated $\xi $-invariant.

        \proclaim{\protag{2.15} {Proposition}}
 Let $X$~be a compact odd dimensional spin manifold and $L\db$~the hermitian
line defined in~\thetag{1.7}.  Then
  $$ \aligned
      L\db&\longrightarrow \Det\db\inv\\
      \{\,\tau (a,T)\in \CC\,\}&\longmapsto \left\{\eta (a) = \tau (a,T)
     \left( \prod\limits_{\lambda >a}\lambda \right)^{1/2}\!\!\! (\Det T)\inv
     \in \bigl(\Det K\db(a) \bigr)\inv \right\}
      \endaligned \tag{2.16} $$
is an isometry.
        \endproclaim

\flushpar
 The proof is straightforward.  First, \thetag{1.5}~and \thetag{1.6}~imply
that $\{\eta (a)\}$~defines an element of~$\Det\inv \db$.  Then
\thetag{1.7}~and \thetag{2.21}~imply that the isomorphism~\thetag{2.16} is an
isometry.  Here, following Ray and Singer~\cite{RS}, we use a $\zeta
$-function to define the infinite product in this isometry.

{}From now on we identify~$L\db$ as the inverse determinant line.  So for {\it
any\/} closed even dimensional spin manifold~$Y$ the hermitian line~$L_Y$ is
defined.

Now we state some properties of the lines~$L_Y$ and the exponentiated $\xi
$-invariant~$\tau _X$.  (It might be illuminating to compare with the
analogous assertions about the Chern-Simons invariant
in~\cite{F2,Theorem~2.19}.)  For simplicity we state these for a single
manifold~$X$ rather than for families.  However, they work as stated for
families, and the proofs are designed to work with the patching construction
of~\S{1}.  (Recall that this is our motivation to allow arbitrary~$a$
in~\thetag{1.2}.)

First, \thetag{2.13}~and \thetag{2.14}~imply that there are isometries
  $$ \align
      L\mstrut _{-Y}&\cong L_Y\inv  \tag{2.17}\\
      L_{Y_1\sqcup Y_2}&\cong L_{Y_1}\gtimes L_{Y_2}. \tag{2.18}\endalign $$
(Note that \thetag{2.17}~is {\it not\/} the inverse of~\thetag{2.13}; the
sign in~\thetag{2.5} enters.  Also, one must keep in mind~\thetag{2.3} when
comparing~\thetag{2.14} and~\thetag{2.18}.)  For the exponentiated $\xi
$-invariant we have
  $$ \align
      \tau \mstrut _{-X} &= \tau _X\inv \\
      \tau _{X_1\sqcup X_2} &= \tau _{X_1}\otimes \tau _{X_2},
      \endalign $$
where we use the isomorphisms~\thetag{2.17} and~\thetag{2.18} to compare the
left and right hand sides of these equalities.

If $Y,Y'$ are spin manifolds, then we define a {\it spin
isometry\/}~$\tilde{f}$ to be an ordinary isometry $f\:Y'\to Y$ together with
a lift $\tilde{f}\:\Spin(Y')\to\Spin(Y)$ to the spin bundle of frames.  A
spin isometry induces an isometry
  $$ L_{Y'}@>\tilde{f}_*>> L_Y  $$
of inverse determinant lines.  If $\tilde{F}\:\Spin(X')\to \Spin(X)$ is a
spin isometry, then
  $$ (\partial \tilde{F})_*(\tau _{X'}) = \tau _X.  $$
Any spin manifold~$Y$ has a naturally defined spin isometry $\tilde{\iota}
\:\Spin(Y)\to \Spin(Y)$ which is multiplication by~$-1\in \Spin_n$; it covers
the identity diffeomorphism on~$Y$.  The induced map on the inverse
determinant line is
  $$ \tilde{\iota}_* = (-1)^{\index D_Y} \tag{2.19} $$

The most important property of the exponentiated $\xi $-invariant is the {\it
gluing law\/}.

        \proclaim{\protag{2.20} {Theorem}}
 Let $X$~be a compact odd dimensional spin manifold, $Y\hookrightarrow X$ a
closed oriented submanifold, and $X\cut$~the manifold obtained by cutting~$X$
along~$Y$.  \rom(See Figure~1.\rom) We assume that the metric on~$X\cut$ is a
product near $\bX\cut = \bX\sqcup Y\sqcup -Y$.  Then
  $$ \tau _X = \Tr_s(\tau _{X\cut}), \tag{2.21} $$
where $\Tr_s$ is the contraction
  $$ L\mstrut _{\bX\cut} @>\thetag{2.18}>> \Lb\mstrut \gtimes L\mstrut
     _Y\gtimes L\mstrut _{-Y} @>\thetag{2.17}>> \Lb\mstrut \gtimes L\mstrut
     _Y\gtimes L_Y\inv @>{\dsize\Tr_s}>> L\db\mstrut \tag{2.22} $$
using the supertrace~\thetag{2.6}.
        \endproclaim


\flushpar
 Notice that $\index D_Y$ is not necessarily zero, which is why we introduce
graded determinant lines.  We prove \theprotag{2.20} {Theorem} in~\S{3}.

To illustrate the gluing law consider an arbitrary closed even dimensional
spin manifold~$Y$ and form the cylinder $C=\moo\times Y$ with the product
metric and spin structure.  Then \break $\tau _C\in L_Y\gtimes L_{-Y}\cong
\End(L_Y)$.  If we cut~$C$ along~$\{0\}\times Y$ we obtain a manifold ``spin
isometric'' to~$C\sqcup C$.  Then \thetag{2.21}~asserts that $\tau _C=\tau
_C\circ \tau _C$, where `$\circ $'~denotes composition in~$\End(L_Y)$.  We
conclude
  $$ \tau _C = \id\in \End(L_Y). \tag{2.23} $$
This equation is derived assuming the gluing law~\thetag{2.21}.  In~\S{4} we
compute it directly ({\theprotag{4.7} {Proposition}}) as part of our proof
of~\thetag{2.21}.

Recall that the circle~$\cir$ admits two inequivalent spin structures, and we
denote the corresponding spin manifolds~`$\cirbdd$' and~`$\cirnonbdd$'.  The
former is the boundary of the disk (with its unique spin structure), while
for the latter the bundle $\Spin(\cirnonbdd)\to SO(\cir)$ is the trivial
double cover of the bundle of oriented orthonormal frames~$SO(\cir)$.  Now
consider $\cirnonbdd\times Y$ with the product metric and product spin
structure.  If we cut along~$\{\pt\}\times Y$ we obtain~$C$, and the gluing
law~\thetag{2.21} asserts
  $$ \tau _{\cirnonbdd\times Y} = \Tr_s(\tau _C) = \Tr_s(\id) = (-1)^{\index
     D_Y}. \tag{2.24} $$
On the other hand, if we apply the spin isometry~$\iota $ to one boundary
component of~$C$ and then glue, we obtain $\cirbdd\times Y$.  It follows
from~\thetag{2.19} that
  $$ \tau _{\cirbdd\times Y} = (-1)^{\index D_Y}\Tr_s(\tau _C) = 1.
     \tag{2.25} $$
Equations~\thetag{2.24} and~\thetag{2.25} agree with known results and
provide a simple check of the signs in the gluing law.

\newpage
\head
\S{3} The Variation formula
\endhead
\comment
lasteqno 3@ 26
\endcomment

The purpose of this section is to present the proof of \theprotag{1.9}
{Theorem}.

Let $\pi \: X \to Z$ be a spin map whose typical fiber is a compact odd
dimensional manifold with boundary. Since the assertion to be proved is
local, it suffices to work over an open set $U_{a,i}$, defined
in~\thetag{1.8}. Over $U_{a,i}$ we have smooth isomorphic hermitian bundles
$K_{\partial X/Z}^{\pm} (a)$ and we choose a smooth family of isometries
  $$ T=T_z(a,i): \ K_{\partial X_z}^+ (a) \ra K_{\partial X_z}^- (a).
     \tag{3.1} $$
By \theprotag{2.15} {Proposition} over the open set $U_{a,i}$, the smooth
section $\tau_{X/Z}$ of $L_{\partial X/Z} \to Z$ can be identified with
  $$ \tau_{X/Z} =  e^{2\pi i \xi_X(a,T)} u^{-1},   $$
where
  $$ u= (\Det T)/\sqrt{\det D_{\partial X/Z}^2 (a)} \in
     \operatorname{Det}_{\partial X/Z}  \tag{3.2} $$
is a smooth section of unit Quillen norm.  Clearly, then, \theprotag{1.9}
{Theorem} is equivalent to the following.

        \proclaim{\protag{3.3} {Theorem}}
  Modulo the integers $\xi_X \bigl(a,T(a,i)\bigr)$ defines a smooth function on
$U_{a,i}$ and
  $$ d\xi_X (a,T) = \comp{\int_{X/Z} \hat{A} (\Omega^{X/Z})}{1} +
     \frac{1}{2\pi i} \,u^{-1} \nabla u. $$
        \endproclaim

As we mentioned earlier the connection $\nabla$ here is the natural unitary
connection on the determinant line bundle introduced in \cite{BF1} by
Bismut-Freed. It is a natural generalization of the induced connection in the
finite dimensional case to the infinite dimensional setting and uses the heat
equation regularization.  For our purpose we recall its construction.
(See~\cite{F1} for a treatment in terms of $\zeta $-functions.)

Let $\pi: \ Y=\partial X \rightarrow Z$ be a spin map and $D^+=D^+_{Y/Z}$ the
family of fiber Dirac operators. (Everything works even if $Y$ is not a
boundary.) Now $D^+$ can be considered as a smooth section of $\Hom (H^+,
H^-)$, where $H^{\pm}$ are infinite dimensional hermitian bundles over $Z$
(see \cite{BF1} for details). Assume for the moment that $H^{\pm}$ are finite
dimensional hermitian bundles over $Z$. In this case the determinant line
bundle can be identified with $(\operatorname{Det} H^-) \gtimes
(\operatorname{Det} H^+)\inv $, and so is naturally endowed with a hermitian
metric. Clearly $\Det D^+$ is a smooth section.  Now if $H^{\pm}$ are also
endowed with unitary connections $\tilde{\nabla}$, then they induce a unitary
connection $\nabla$ on the determinant line bundle.  In fact when $D^+$ is
invertible,
  $$ \nabla \Det D^+ = \operatorname{Tr}[(D^+)^{-1}\tilde{\nabla} D^+]\cdot
     \Det D^+ .  \tag{3.4} $$
Further if $H^{\pm}=K^{\pm} \oplus H^{\pm}_1$ is an orthogonal decomposition
invariant under $D^+$, then
  $$ \nabla=\nabla^K + \nabla^{H_1}.   \tag{3.5} $$
These two properties fully suggest how to define it in the infinite
dimensional setting.

Thus over $U_a$ let
  $$ H^{\pm}=K^{\pm}(a) \oplus H^{\pm}(a) $$
be the orthogonal decomposition defined in~\S1. The infinite dimensional
hermitian bundles $H^{\pm}$ are equipped with the unitary connection
$\tilde{\nabla}$ defined in \cite{BF2, Def. 1.3}.\footnote{Note that the
notation there for that connection is~`$\tilde\nabla ^u$'.} Over
$U_a$ we have smooth finite dimensional subbundles $K^{\pm}(a)$ of $H^{\pm}$.
Hence they inherit a unitary connection, which in turn induces a unitary
connection $\nabla^a$ on $(\operatorname{Det} K^-(a))\gtimes
(\operatorname{Det} K^+(a))\inv $. By the additivity \thetag{3.5} this is the
$K^{\pm}(a)$-part of the connection.

To define the $H^{\pm}(a)$-part of the connection one makes sense of
\thetag{3.4} in the infinite dimensional setting by the heat equation
regularization.  Note that the restriction $D^+(a)$ of $D^+$ to $H^+(a)$ is
indeed invertible. When there is no confusion we also use `$D^2(a)$' (instead
of `$D^-(a)D^+(a)$') to denote the restriction of $D^2$ to $H^+(a)$.  The
formal expression $\operatorname{Tr}[(D^+(a))^{-1}\tilde{\nabla} D^+(a)]$
will be defined by
  $$ \operatorname{Tr}[(D^+(a))^{-1}\tilde{\nabla}
     D^+(a)]=\operatorname{f.p.}
     \Big\{\operatorname{Tr}[(D^+(a))^{-1}\tilde{\nabla}
     D^+(a)e^{-tD^2(a)}]\Big\}, \tag{3.6} $$
where f.p. is a suitably defined finite part of the right hand side of
\thetag{3.6} as $t \to 0$.

To define this finite part, note that
  $$ \operatorname{Tr}[(D^+(a))^{-1}\tilde{\nabla}
     D^+(a)e^{-tD^2(a)}]=\int_t^{\infty}
     \operatorname{Tr}[(D^-(a))\tilde{\nabla} D^+(a)e^{-sD^2(a)}] \,ds.
       $$
It follows that as $t \to 0$
  $$ \operatorname{Tr}[(D^+(a))^{-1}\tilde{\nabla} D^+(a)e^{-tD^2(a)}] \sim
     \sum_{j=-n/2}^{-1} a_j t^j +a_0 +a_{0,1} \log t + \sum_{j=1}^{\infty}
     a_j t^j .    $$
Then the finite part is defined as
  $$ \operatorname{f.p.}\Big\{\operatorname{Tr}[(D^+(a))^{-1}\tilde{\nabla}
     D^+(a)e^{-tD^2(a)}]\Big\} =a_0 + \Gamma'(1) a_{0,1}, $$
or symbolically,
  $$ \split
      \operatorname{f.p.}
     \Big\{\operatorname{Tr}[(D^+(a))^{-1}\tilde{\nabla}
     D^+(a)e^{-tD^2(a)}]\Big\}=&\LIM_{t \to 0}
     \operatorname{Tr}[(D^+(a))^{-1}\tilde{\nabla} D^+(a)e^{-tD^2(a)}]  \\
      &\qquad + \Gamma'(1) \LIM_{t \to 0} \frac{1}{\log t}
     \operatorname{Tr}[(D^+(a))^{-1}\tilde{\nabla} D^+(a)e^{-tD^2(a)}].
      \endsplit \tag{3.7} $$
Finally the Bismut-Freed connection is defined as
  $$ \nabla =\nabla^a +
     \operatorname{f.p.}\Big\{\operatorname{Tr}[(D^+(a))^{-1}\tilde{\nabla}
     D^+(a)e^{-tD^2(a)}]\Big\}.  $$

Coming back to \theprotag{3.3} {Theorem}, when $D_{\partial X/Z}$ is
invertible we can choose $a$ less than the smallest nonzero eigenvalues of
$D_{\partial X/Z}$. In this case $u=\frac{\Det D_{\partial X/Z}^+}{\|\det
D_{\partial X/Z}^+ \|}$ and thus $u^{-1} \nabla u = \operatorname{Im}\,
\omega$, where $\omega$ is the connection form for the Bismut-Freed
connection:
  $$ \nabla (\Det D_{\partial X/Z}^+) = \omega\cdot \Det D_{\partial
     X/Z}^+, $$
The imaginary part of $\omega$ has the following explicit formula:
  $$ \operatorname{Im}\,\omega = \frac{1}{2} \int_0^\infty
     \operatorname{Tr}_s (D_{\partial X/Z} \tilde{\nabla} D_{\partial X/Z}
     e^{-t D_{\partial X/Z}^2}) \,dt .   \tag{3.8} $$
That the integral in \thetag{3.8} is well-defined comes from the following
cancellation result (\cite{BF2}, \cite{C2}):
  $$ \operatorname{Tr}_s (D_{\partial X/Z} \tilde{\nabla} D_{\partial X/Z}
     e^{-t D_{\partial X/Z}^2}) = O(1) \qquad \text{as $t\to0$.}
      \tag{3.9} $$
This result holds without the assumption on the invertibility of $D_{\partial
X/Z}$ and is also crucial in our proof of \theprotag{3.3} {Theorem}.

Thus in the invertible case our formula states
  $$ \aligned
      d\xi_X &= \comp{\int_{X/Z} \hat{A} (\Omega^{X/Z})}{1} + \frac{1}{4\pi
     i}\int_0^\infty \operatorname{Tr}_s (\tilde{\nabla} D_{\partial X/Z}
     \cdot D_{\partial X/Z} e^{-t D_{\partial X/Z}^2}) \,dt \\
      &= \comp{\int_{X/Z} \hat{A} (\Omega^{X/Z}) - \tilde{\eta}}{1},
     \endaligned  $$
where $\tilde{\eta}$ is the differential form generalization of $\eta$
introduced in \cite{BC2}. We point out that Cheeger~\cite{C2} has also proven a
formula similar to the above in the context of conical singularity.

The proof of \theprotag{3.3} {Theorem} is divided into several lemmas and two
propositions.

 The first lemma deals with a special case.  Namely, we assume that the
metrics along the fibers are of the form
  $$  g_z = du^2 + g_{\partial X_z}, $$
near the boundary, where $g_{\partial X_z}$ is independent of $z$, i.e. the
metrics near the boundary are all the same (and of product type). Fix a choice
of boundary condition $\langle a,T  \rangle$.

        \proclaim{\protag{3.10} {Lemma}}
 Under these conditions $\xi(a,T) \pmod1$ is a smooth function on $U_a$ and
  $$  d \xi (a,T) = -\frac{1}{\sqrt{\pi}}\LIM_{t \to 0} t^{1/2}
     \operatorname{Tr} (\tilde{\nabla}D(a,T) e^{-tD^2 (a,T)}), $$
 where {\rm LIM} means taking the constant term in the asymptotic expansion.
         \endproclaim

        \demo{Proof}
 This is a slight generalization of \cite{M\"u,Prop. 2.15}. His proof can be
easily generalized to this situation and is given in \theprotag{A.17}
{Proposition}.
        \enddemo

In general the boundary geometry and the boundary conditions vary.  The idea
here is to conjugate to a family with fixed boundary conditions.

Thus write the metric  $g_z$ near the boundary as
  $$  g_z = du^2 + g_{\partial X_z}. $$
and let $\Pi_a(z)$ denote the orthogonal projection onto the space spanned by
eigensections of $D_{\partial M} (z)$ with eigenvalues $\lambda > \sqrt{a}$.
Then $\Pi_a(z)$ is a smooth family of (pseudodifferential) projections on
$L^2(\partial X/Z,S)$ (for $z\in U_a$), and let $\Pi_T(z)$ denote the
corresponding orthogonal projection onto the graph of $T_z(a,i)$, defined
in~\thetag{3.1}. Then
  $$  \Pi_{(a,T)}(z) = \Pi_a(z) + \Pi_T(z) $$
is a smooth family of pseudodifferential projections which describes the
family of the boundary conditions. From the general perturbation theory, for
any fixed $z_0 \in Z$ there is a smooth family of unitary operator $U(z)$ on
$L^2(\partial X_z,S)$ (see \cite{D2, Lemma 2.9}, for example) such that
   $$ \aligned
       U(z) \Pi_{(a,T)}(z_0) U^{-1}(z)  &= \Pi_{(a,T)} (z) \\
       U(z_0) &= Id.
     \endaligned  $$
In fact, as we will see later,
  $$ U(z) = \pmatrix B^{-1}(z)B(z_0) & 0 \\ 0 & 1 \endpmatrix , \tag{3.11} $$
where $B(z) = T(z) \oplus \frac{D_{\partial X_z} (a)}{\sqrt{D_{\partial
X_z}^2 (a)}} : H^+ \to H^-$.

Now extend $U(z)$ to a smooth family of unitary operators on $L^2(X/Z,S)$
such that $U(z)$ is constant along normal directions to a neighborhood of
$\partial X/Z$ and identity in the interior and interpolate in between. This
can be done, at least in a neighborhood of $z_0$. For example, let $\chi(u)$
be a smooth function on $[0, \ 1]$ such that $\chi(u) =0$ for $u\geq 3/4$ and
$\chi(u) =1$ for $u\leq 1/2$.  Then $U\bigl(\chi(u)z+(1-\chi(u))z_0\bigr)$
does the job.  (Here we interpret $z$ as local coordinates around $z_0$.)
For simplicity we still denote this extension by $U(z)$.

        \proclaim{\protag{3.12} {Lemma}}
 Modulo the integers  $\xi(a,T(z))$ defines a smooth function on $U_a$ and
  $$ \split
       d \xi (a,T(z)) = &-\frac{1}{\sqrt{\pi}}\LIM_{t \to 0} t^{1/2}
     \operatorname{Tr} (\tilde{\nabla}D(a,T) e^{-tD^2 (a,T)})
       \\
       &\qquad - \frac{1}{\sqrt{\pi}}\LIM_{t \to 0} t^{1/2}
     \operatorname{Tr}[D(a,T), \tilde{\nabla} U e^{-tD^2(a,T)}].  \endsplit
     \tag{3.13} $$
        \endproclaim

        \demo{Proof}
  Since $D(a,T(z))$ and $U(z)^{-1}D(a,T(z))U(z)$ have the same eigenvalues we
have
  $$ \xi(a,T) = \xi \bigl(U(z)^{-1}D(a,T(z))U(z)\bigr). $$
But now $U(z)^{-1}D(a,T(z))U(z)$ is a smooth family of operators satisfying
conditions~(Ha), (Hb), and (Hc), which are defined in the Appendix preceding
\theprotag{A.14} {Lemma} and \theprotag{A.15} {Lemma}.  Therefore, we apply
\theprotag{A.17} {Lemma} of the Appendix to obtain
  $$ \align
       d \xi \bigl(U(z)^{-1}D(a,T(z))&U(z)\bigr) \\
       &= -\frac{1}{\sqrt{\pi}}\LIM_{t \to 0} t^{1/2} \operatorname{Tr}
     (\tilde{\nabla}[U(z)^{-1}D(a,T(z))U(z)]e^{-t
     (U(z)^{-1}D(a,T(z))U(z))^2}) \\
      \split &= -\frac{1}{\sqrt{\pi}}\LIM_{t \to 0} t^{1/2} \operatorname{Tr}
     (\tilde{\nabla}D(a,T) e^{-tD^2 (a,T)}) \\
      &\qquad \qquad -\frac{1}{\sqrt{\pi}}\LIM_{t \to 0} t^{1/2}
     \operatorname{Tr}[D(a,T), \tilde{\nabla} U e^{-tD^2(a,T)}] .
     \endsplit\endalign $$
        \enddemo

        \rem
 In the second term of~\thetag{3.13}, $[D(a,T), \tilde{\nabla} U
e^{-tD^2(a,T)}]$ should be interpreted as an operator acting on the Sobolev
space~$H^1(X,S)$.  As we see from the proof, this term comes from
$[D(a,T),\tilde{\nabla}U]e^{-tD^2(a,T)}$, which is clearly trace class
on~$L^2(X,S)$.  Of course, both traces are equal.
        \endrem

We now look at the first term in \thetag{3.13}.

        \proclaim{\protag{3.14} {Proposition}}
 We have
  $$ -\frac{1}{\sqrt{\pi}}\LIM_{t \to 0} t^{1/2}
     \operatorname{Tr} (\tilde{\nabla}D(a,T) e^{-tD^2 (a,T)})=
\comp{\int_{X/Z}
     \hat{A}(\Omega^{X/Z})}{1}. $$
        \endproclaim

        \demo{Proof}
 By the explicit construction of the heat kernel $e^{-tD^2(a,T)}$
(see~\thetag{A.8}), the asymptotic expansion separates into an interior
part and a boundary part, and by the corresponding result for closed manifold
we have
  $$ -\frac{1}{\sqrt{\pi}}\LIM_{t \to 0} t^{1/2}
     \operatorname{Tr} (\tilde{\nabla}D(a,T) e^{-tD^2 (a,T)})=
\comp{\int_{X/Z}
     \hat{A}(\Omega^{X/Z})}{1} + \text{boundary term}. $$
As to computing the boundary term we can replace the manifold $X/Z$ by the
half cylinder ${\Bbb R}_+\times {\partial X/Z}$, with the family of the
metrics given by
  $$ g_z=du^2+g_{\partial X_z}. $$
To compute the heat kernel $e^{-t D^2(a,T)}$ on the half cylinder, we let $\{
\varphi_{\lambda} \}$ be an orthonormal basis of eigensections of
$D_{\partial X/Z}$ such that $J\varphi_{\lambda}=\varphi_{-\lambda}$. Then
  $$ e^{-t D^2(a,T)} = E_{>a}(t) + E_{<a}(t),  \tag{3.15} $$
where
  $$ \split
       E_{>a}(t) = &\sum_{\lambda > \sqrt{a}} \frac{e^{-\lambda^2t}}{\sqrt{4
     \pi t}}(e^{-(u-v)^2/4t}-e^{-(u+v)^2/4t}) \,\varphi_{\lambda}\otimes
     \varphi_{\lambda}^* \\
       &\qquad + \frac{e^{-\lambda^2t}}{\sqrt{4 \pi
     t}}(e^{-(u-v)^2/4t}+e^{-(u+v)^2/4t}) \,J\varphi_{\lambda}\otimes
     J\varphi_{\lambda}^* \\
      &\qquad - \lambda
     e^{\lambda(u+v)}\operatorname{erfc}(\frac{u+v}{2\sqrt{t}}+ \lambda
     \sqrt{t}) \,J\varphi_{\lambda}\otimes J\varphi_{\lambda}^*, \endsplit $$
with
  $$ \operatorname{erfc}(x)=\frac{2}{\sqrt{\pi}} \int_{x}^{\infty} e^{-\xi^2}
     d\xi, $$
and $E_{<a}(t)$ is the heat kernel of the following  system  on the half
line~$u\ge0$:
  $$ \left\{ \aligned (\partial_t - \partial_u^2+A^2) E_{<a}(t,u,v) &= 0 \\
     E_{<a}|_{t=0}&= \operatorname{Id} \\ \Pi_TE_{<a}|_{u=0} &=0\\ \ J\Pi_T
     J(\partial_u + A)E_{<a}|_{u=0} &= 0.  \endaligned \right. $$
Here $A = D_{\partial X/Z}|_{K(a)}$. Note that $A$ is a smooth family of
finite dimensional (symmetric) endomorphisms and the boundary condition here
is local.

Therefore,
  $$ \split
      \operatorname{tr}(\tilde{\nabla}D(a,T)E_{>a}(t))(u) &= \sum_{\lambda >
     \sqrt{a}} \frac{e^{-\lambda^2t}}{\sqrt{4 \pi t}}(1-e^{-u^2/t}) \,\langle
     J\tilde{\nabla}D\varphi_{\lambda}, \ \varphi_{\lambda}\rangle \\
      &\qquad \qquad + \frac{e^{-\lambda^2t}}{\sqrt{4 \pi
     t}}(1+e^{-u^2/t})\,\langle \tilde{\nabla}D\varphi_{\lambda}, \
     J\varphi_{\lambda}\rangle \\
      &\qquad\qquad -\lambda e^{2\lambda
     u}\operatorname{erfc}(\frac{u}{\sqrt{t}}+ \lambda \sqrt{t}) \,\langle
     \tilde{\nabla}D\varphi_{\lambda}, \ J\varphi_{\lambda}\rangle \\
      &= \sum_{\lambda >\sqrt{a}} \frac{d}{du} \[\frac{1}{2} e^{2\lambda u}
     \operatorname{erfc} (\frac{u}{\sqrt{t}} + \lambda \sqrt{t})\] \,\langle
     J\tilde{\nabla}D \varphi_\lambda, \varphi_\lambda \rangle.  \endsplit $$
Here, and also in what follows, we have suppressed the subscript $\partial
X/Z$ of $D$. Integrating with respect to $u$ from $0$ to $\infty$ yields
  $$ \aligned
     \operatorname{Tr}(\tilde{\nabla}D(a,T)E_{>a}(t))  &=  \sum_{\lambda >
     \sqrt{a}} \frac{1}{2}\operatorname{erfc} (\lambda \sqrt{t})\langle
     J\tilde{\nabla}D \varphi_\lambda, \varphi_\lambda \rangle \\  &=
     \frac{i}{2\sqrt{\pi}} \int_{\sqrt{t}}^\infty
     \operatorname{Tr}_s(D(a)\tilde{\nabla}D(a) e^{-s^2 D^2(a)})\, ds.
     \endaligned  $$
Here the last equation follows from the fact that
  $$ \langle J\tilde{\nabla}D \varphi_{-\lambda}, \varphi_{-\lambda} \rangle
     = -\langle J\tilde{\nabla}D \varphi_\lambda, \varphi_\lambda \rangle $$
which is a consequence of the following equations:
  $$ J\tilde{\nabla}D=-\tilde{\nabla}D J, \qquad
     J\varphi_\lambda=\varphi_{-\lambda}. \tag{3.16} $$
Now,
  $$ \operatorname{Tr}_s\bigl(D(a)\tilde{\nabla}D(a) e^{-s^2 D^2(a)}\bigr) =
     O(1) \qquad \text{as $t\to0$}, $$
as it follows from \thetag{3.9}. Consequently,
  $$ \LIM_{t \to 0} t^{1/2}
     \operatorname{Tr}\bigl(\tilde{\nabla}D(a,T)E_{>a}(t)\bigr) = 0. $$
On the other hand,
  $$ \operatorname{Tr}\bigl(\tilde{\nabla}D(a,T)E_{<a}(t)\bigr) =
     \operatorname{Tr} \bigl(J\tilde{\nabla}D E_{<a}(t)\bigr) =
     i \operatorname{Tr}_s\bigl(\tilde{\nabla}D E_{<a}(t)\bigr) . $$
By \thetag{3.16} $\tilde{\nabla}D$ is an odd operator. However the heat kernel
$E_{<a}(t)$ is not even because of the boundary condition. The crucial
observation here is that the leading asymptotic as~$t\to0$ is indeed even,
for local boundary conditions do not contribute to the leading asymptotic.
Since the underlying manifold here is one dimensional, the leading asymptotic
is $t^{-1/2}$, which implies
  $$ \operatorname{Tr}_s( \tilde{\nabla}D E_{<a}(t)) =O(1)\
     \operatorname{as}\ t \to 0. $$
Therefore,
  $$ \LIM_{t \to 0} t^{1/2}
     \operatorname{Tr}\bigl(\tilde{\nabla}D(a,T)E_{<a}(t)\bigr) = 0. $$
Thus the boundary  term is zero. This finishes our proof.
        \enddemo

We now turn to the computation of the commutator term in \thetag{3.13}. In
general the trace of the commutator of a bounded linear operator with a trace
class operator is zero. On a closed manifold, this can be extended to
  $$  \operatorname{Tr}\,[D,K] = 0 $$
for $D$ a differential operator and $K$ a smoothing operator (say). This is
no longer true on a manifold with boundary. However, we have

        \proclaim{\protag{3.17} {Lemma}}
 For $D$ the Dirac operator and $K$ a smoothing operator with smooth kernel
$K(x,x')$ on a compact spin manifold~$M$ with boundary we have
  $$ \operatorname{Tr} [D,K] = -\int_{\partial M} \operatorname{tr}
     \bigl(JK(y,y)\bigr) \,d \operatorname{vol} (y).  \tag{3.18} $$
        \endproclaim

        \rem
 This is quite similar to the characteristic feature of the b-trace
introduced by Melrose \cite{M} in the context of manifolds with asymptotically
cylindrical ends.
        \endrem

        \demo{Proof}
 Clearly $DK$ is a smoothing operator with kernel given by $D_x K(x,x')$. Thus
  $$ \operatorname{Tr} (DK) = \int_M \operatorname{tr} (D_x K(x,x')|_{x'=x})\,
     d vol(x). $$
On the other hand,
  $$ \split
       (KD) f(x)  &=  \int_M K(x,x') (Df)(x') d vol (x') \\ &=  \int_M
     D_{x'} K(x,x') f(x') d vol (x') + \int_{\partial M} JK(x,y') f(y') d vol
     (y'). \endsplit  $$
Therefore the kernel of $KD$ is given by $D_{x'}K(x,x') +
JK(x,x') \delta_{\partial M}$. And hence
  $$ \split
       \operatorname{Tr} [D,K]  &=  \operatorname{Tr} (DK)
     -\operatorname{Tr} (\overline{(DK)^*}) - \int_M \operatorname{tr} JK(x,x)
     \delta_{\partial M} \,dvol(x) \\  &=  -\int_{\partial M}
     \operatorname{tr} (JK(y,y)) \,dvol(y).  \endsplit $$

It should be pointed out that for the above equation the Lidskii's theorem
does not apply immediately to $JK(x, x')\delta_{\partial M}$. But this can be
overcome by approximating the delta function via smooth functions and
estimating the trace norm of the approximation via the Hilbert-Schmidt norms.
        \enddemo

With this lemma at our disposal we now turn to the commutator term.  Recall
the definition of~$u$ from~\thetag{3.2}.

        \proclaim{\protag{3.19} {Proposition}}
 We have
  $$ \LIM_{t \to 0} t^{1/2} \operatorname{Tr} [D(a,T),
     \tilde{\nabla}U e^{-tD^2(a,T)}] = \frac{i}{2\sqrt{\pi}} \,u^{-1} \nabla
u.
       $$
        \endproclaim

        \demo{Proof}
 Clearly $\tilde{\nabla}U e^{-tD^2(a,T)}$ is a smoothing operator. Therefore
according to \thetag{3.18} the trace of the commutators part can be
computed by taking pointwise trace of the Schwartz kernel of $\tilde{\nabla}U
e^{-tD^2(a,T)}$ and integrated over the boundary. Thus $U$ can be taken to be
the original family of unitary operators on the boundary, extended radially
constant to the whole cylinder. For our computation we need the precise
construction of $U$.

Recall that $U$ is constructed to conjugate the family of boundary
conditions, which are described by (see \thetag{1.2})
  $$ W_{\langle a,T\rangle} = \{ \langle \phi^+, \phi^- \rangle \in
     H_{\partial X/Z}: \phi^- + (T \oplus \frac{D_{\partial X/Z}
     (a)}{\sqrt{D_{\partial X/Z}^2 (a)}}) \phi^+ = 0 \}. $$
In other words, they are described by the graph of the pseudodifferential
operator:
  $$  B(z) = T(z) \oplus \frac{D_{\partial X_z} (a)}{\sqrt{D_{\partial X_z}^2
     (a)}} : H_{\partial X_z}^+ \ra H_{\partial X_z}^-. $$
Then it is not hard to verify that formula \thetag{3.11} defines such a
unitary conjugation. One easily finds
  $$ \tilde{\nabla}U (z_0) = \pmatrix -B^{-1}(z_0)\tilde{\nabla} B(z_0) & 0
     \\ 0 & 0 \endpmatrix $$
and
  $$ B^{-1}\tilde{\nabla}B = T^{-1}\tilde{\nabla}T \oplus \bigl((D^+ (a))^{-1}
     \tilde{\nabla}D^+(a) -\frac{1}{2} (D^2 (a))^{-1} \tilde{\nabla}(D^2
     (a)) \bigl). $$
Using these and \thetag{3.15} we obtain
  $$ \split
      -\int_{\partial X/Z} \operatorname{tr} J \tilde{\nabla}U e^{-tD^2(a,T)}
     = &\operatorname{tr} (JT^{-1}\tilde{\nabla}T E_{<a}(t) |_{u=0} ) \\
      &\quad + \int_{\partial X/Z} \operatorname{tr} (J[(D^+ (a))^{-1}
     \tilde{\nabla}D^+(a) -\frac{1}{2} (D^2 (a))^{-1}\tilde{\nabla} (D^2
     (a))] E_{>a}(t)).  \endsplit \tag{3.20} $$
For the first term we have
   $$ \split
      \LIM_{t \to 0} t^{1/2}\operatorname{tr}
     (JT^{-1}\tilde{\nabla}T E_{<a}(t) |_{u=0} )  &=  \frac{1}{\sqrt{4\pi}}
     \operatorname{tr} (JT^{-1}\tilde{\nabla}T) \\  &= \frac{i}{2
     \sqrt{\pi}} \frac{\nabla^a\Det T}{\Det T}, \endsplit
     \tag{3.21} $$ 
where again we have made use of the observation that the leading asymptotic
of  $\operatorname{tr} (JT^{-1}\tilde{\nabla}T E_{<a}(t))$ is
independent of the boundary condition.

The second term is a little bit more complicated. We first note that
  $$ E_{>a}(t) |_{\partial X/Z} = \sum_{\lambda > \sqrt{a}}
     (\frac{e^{-\lambda^2t}}{\sqrt{\pi t}} - \lambda \operatorname{erfc}
     (\lambda \sqrt{t})) \,J\varphi_\lambda \otimes J\varphi_\lambda^*, $$
and
  $$ \split
      \lambda \operatorname{erfc} (\lambda \sqrt{t}) &=
     \frac{2\lambda^2}{\sqrt{\pi}} \int_t^\infty \frac{1}{2\sqrt{s}} e^{-s
     \lambda^2} \,ds \\ &= \frac{1}{\sqrt{\pi t}} e^{-t\lambda^2} -
     \frac{1}{2\sqrt{\pi}} \int_t^\infty s^{-3/2}e^{-s \lambda^2} \,ds.
     \endsplit  $$
Hence
  $$  E_{>a}(t) \res{\partial X/Z} = \sum_{\lambda > \sqrt{a}}
     \frac{1}{2\sqrt{\pi}} \int_t^\infty s^{-3/2}e^{-s \lambda^2}\,
     dsJ\varphi_\lambda \otimes J\varphi_\lambda^*, $$
and
  $$ \split
      \int_{\partial X/Z} \operatorname{tr} \bigl(J[(D^+ (a))^{-1}
     \tilde{\nabla} &D^+(a) -\frac{1}{2} ((D^2 (a))^{-1}\tilde{\nabla} (D^2
     (a))] E_{>a}(t)\bigr)\\
      &= \frac{i}{4\sqrt{\pi}} \int_t^\infty s^{-3/2} \operatorname{Tr}
     \bigl((D^+ (a))^{-1} \tilde{\nabla} D^+(a) e^{-s D^2 (a)}\bigr) \,ds \\
      &\qquad - \frac{i}{8\sqrt{\pi}} \int_t^\infty s^{-3/2}
     \operatorname{Tr} \bigl((D^2 (a))^{-1} \tilde{\nabla} (D^2(a)) e^{-s D^2
     (a)}\bigr) \,ds.  \endsplit $$
One finds
  $$ \multline
      \LIM_{t \to 0} t^{1/2}\int_{\partial X/Z} \operatorname{tr}\bigl(J[
     (D^+ (a))^{-1} \tilde{\nabla} D^+(a) -\frac{1}{2} ((D^2
     (a))^{-1}\tilde{\nabla} (D^2 (a))] E_{>a}(t)\bigr) \\
      =\frac{i}{2\sqrt{\pi}}\LIM_{t \to 0} \operatorname{Tr} \bigl((D^+
     (a))^{-1} \tilde{\nabla} D^+(a) e^{-t D^2 (a)}\bigr) +
     \frac{i}{\sqrt{\pi}}\LIM_{t \to 0} \frac{1}{\log t} \operatorname{Tr}
     \bigl((D^+(a))^{-1} \tilde{\nabla} D^+(a) e^{-t D^2 (a)}\bigr)\\
      - \frac{i}{4\sqrt{\pi}}\LIM_{t \to 0}\operatorname{Tr}
     \bigl((D^2 (a))^{-1} \tilde{\nabla} (D^2(a)) e^{-t D^2 (a)}\bigr) -
     \frac{i}{2\sqrt{\pi}}\LIM_{t \to 0} \frac{1}{\log t} \operatorname{Tr}
     \bigl((D^2 (a))^{-1} \tilde{\nabla} (D^2(a)) e^{-t D^2 (a)}\bigr).
     \endmultline \tag{3.22} $$ 
{}From~\thetag{3.9} and the identity
 $$ \operatorname{Tr}_s[(D(a))^{-1}\tilde{\nabla}
     D(a)e^{-tD^2(a)}]=\int_t^{\infty}
     \operatorname{Tr}_s[(D(a))\tilde{\nabla} D(a)e^{-sD^2(a)}] \,ds.
\tag{3.23} $$
we find
 $$ \LIM_{t \to 0}  \frac{1}{\log t} \operatorname{Tr}_s[(D(a))^{-1}
\tilde{\nabla} D(a)e^{-tD^2(a)}] =0, $$
or equivalently,
  $$ \LIM_{t \to 0} \frac{1}{\log t} \operatorname{Tr} \bigl((D^+ (a))^{-1}
     \tilde{\nabla} D^+(a) e^{-t D^2 (a)}\bigr) = \frac{1}{2} \LIM_{t \to 0}
     \frac{1}{\log t} \operatorname{Tr} \bigl((D^2 (a))^{-1} \tilde{\nabla}
     D^2(a) e^{-t D^2 (a)}\bigr). \tag{3.24} $$
Thus the right hand side of~\thetag{3.22} reduces to
  $$ \frac{i}{2\sqrt{\pi}}\LIM_{t \to 0} \operatorname{Tr} \bigl((D^+
     (a))^{-1} \tilde{\nabla} D^+(a) e^{-t D^2 (a)}\bigr) -
 \frac{i}{4\sqrt{\pi}}\LIM_{t \to 0}\operatorname{Tr}
     \bigl((D^2 (a))^{-1} \tilde{\nabla} (D^2(a)) e^{-t D^2 (a)}\bigr). $$

On the other hand, we have by~\thetag{3.7}
  $$ \split
      \frac{\nabla \Det T}{\Det T}= \frac{\nabla^a\Det T}{\Det T} &+ \LIM_{t
     \to 0} \operatorname{Tr} \bigl((D^+ (a))^{-1} \tilde{\nabla} D^+(a)
     e^{-t D^2 (a)}\bigr) \\
      &\qquad + \Gamma'(1) \LIM_{t \to 0} \frac{1}{\log t} \operatorname{Tr}
     \bigl((D^+ (a))^{-1} \tilde{\nabla} D^+(a) e^{-t D^2 (a)}\bigr)
     \endsplit \tag{3.25} $$ 
and
  $$ \split
      \frac{d (\sqrt{\det D^2 (a)})}{\sqrt{\det D^2 (a)}} =
     \frac{1}{2}\LIM_{t \to 0}&\operatorname{Tr} \bigl((D^2 (a))^{-1}
     \tilde{\nabla}(D^2(a)) e^{-t D^2 (a)}\bigr) \\
      &\qquad +\frac{1}{2}\Gamma'(1) \LIM_{t \to 0} \frac{1}{\log t}
     \operatorname{Tr} \bigl((D^2 (a))^{-1} \tilde{\nabla} D^2(a) e^{-t D^2
     (a)}\bigr) .  \endsplit \tag{3.26} $$ 
We combine \thetag{3.20}--\thetag{3.26} to complete the proof.
        \enddemo

\newpage
\head
\S{4} The Gluing Formula
\endhead
\comment
lasteqno 4@ 23
\endcomment

In this section we prove \theprotag{2.20} {Theorem}.  We assume the notation
of that theorem and of~\S{1}.

Fix a positive number $a'\notin \spec(D\db^2)$.  Choose an isometry
  $$ T'\:K\db^+(a')\to K\db^-(a').  $$
Then according to~\thetag{1.7} and~\thetag{2.16}, the pair~$\langle a',T'
\rangle$ induces a trivialization of~$L\db$.  This trivialization is simply
carried along in the computation below.  Much more essential is the
following.  Choose~$a\notin\spec(D_Y^2)$ and denote
  $$ K^\pm = K_Y^\pm(a) = K^\mp_{-Y}(a).   $$
Now choose an isometry
  $$ T\: K^+\oplus K^-\longrightarrow K^+\oplus K^-. \tag{4.1} $$
Note that $T$~has a numerical determinant~$\det T\in \CC$.  Now $K^+_{Y\sqcup
-Y}\cong K^+\oplus K^-$ and $K^-_{Y\sqcup -Y}\cong K^-\oplus K^+$ (note the
swap in factors from the right hand side of~\thetag{4.1}), so there is an
induced trivialization
  $$ (-1)^{\dim K^+  \dim K^-}(\Det T)\inv \in L_{Y\sqcup -Y}. \tag{4.2}
     $$
Our first task is to compute the image of~\thetag{4.2} under the sequence of
maps~\thetag{2.22}, where we leave off the $L\db$~factor for convenience.
Recall that \thetag{2.22}~is the composition
  $$ \Tr_s\circ \thetag{2.17}\circ \thetag{2.18}. \tag{4.3} $$
Each of the three maps in~\thetag{4.3} involves a factor, and these factors
are computed in~\thetag{2.9}, \thetag{2.10}, and~\thetag{2.11}.  The total
factor (including the factor in~\thetag{4.2}) is
  $$ (-1)^{\dim K^+  \dim K^-} (-1)^{\dim K^+ + \dim K^-}
     (-1)^{\dim K^+} (-1)^{\dim K^+(\dim K^+ + \dim K^-)}=\sind, $$
from which it follows that the image of~\thetag{4.2} is
  $$ (-1)^{\dim K^+  \dim K^-}(\Det T)\inv @>\thetag{2.22}>> \sind(\det
     T)\inv . \tag{4.4} $$
Thus equation~\thetag{2.21} is equivalent to the following statement.

        \proclaim{\protag{4.5} {Proposition}}
 Let $X$~be a compact odd dimensional spin manifold, $Y\hookrightarrow X$ a
closed oriented submanifold, and $X\cut$~the manifold obtained by cutting~$X$
along~$Y$.  We assume that the metric on~$X\cut$ is a product near $\bX\cut =
\bX\sqcup Y\sqcup -Y$.  Choose~$a,a',T,T'$ as above.  Then
  $$ \tau _{X\cut}(a',T';a,T) = \sind\det T\cdot \tau _X(a',T'). \tag{4.6} $$
        \endproclaim

\flushpar
 Equation~\thetag{4.6} is an equality of complex numbers.

As a preliminary to proving \theprotag{4.5} {Proposition} we compute directly
the exponentiated $\xi $-invariant of the cylinder.  This
generalizes~\cite{LW,\S3}.

        \proclaim{\protag{4.7} {Proposition}}
 Let $Y$~be a closed even dimensional spin manifold and $C=\moo\times Y$ the
corresponding cylinder.  Choose~$a,T$ as above to define boundary conditions
for the Dirac operator on~$C$.  Then
  $$ \tau _C(a,T) = \det T. \tag{4.8} $$
        \endproclaim

\flushpar
 This is compatible with~\thetag{2.23}, which we derived in~\S{2} as a
consequence of the gluing law.\footnote{Of course, that derivation was not a
proof as the proof of the gluing law depends on \theprotag{4.7}
{Proposition}.} Namely, the element of~$\End(L_Y)$ corresponding
to~\thetag{4.8} is~$\tau _C(a,T)(\Det T)\inv $---the $\zeta $-factor
in~\thetag{2.19} cancels out for~$\End(L_Y)$---and as in~\thetag{4.4} we
compute
  $$ \tau _C(a,T)\bigl((-1)^{\dim K^+ \dim K^-}(\Det T)\inv \bigr)
     @>\thetag{2.22}>> \sind \tau _C(a,T)(\det T)\inv = \sind, \tag{4.9} $$
which agrees with the supertrace of~$\id\in \End(L_Y)$.

        \demo{Proof}
 We first prove~\thetag{4.8} assuming that $a=\epsilon $~is less than the
first positive eigenvalue of~$D_Y^2$.  In other words, $K=K^+\oplus K^-$ is
the kernel of~$D_Y$.  Then we use the variation formulas of~\S{3} to derive
the general formula.

A spinor field on~$C$ is a sum of fields of the form
  $$ \psi =f(t)\phi ^+_\lambda  + g(t)\phi ^-_{\lambda }, \tag{4.10} $$
where $f,g\:\moo\to\CC$ and $\phi ^\pm_\lambda \in E^\pm_Y(\lambda )$ are
eigenfunctions of~$D_Y^2$.  If~$\lambda >0$ we choose $\phi ^-_\lambda
=D_Y\phi _\lambda ^+$, and then
  $$ D_C\psi =\bigl(-if'(t) +i\lambda g(t) \bigr)\phi ^+_\lambda  +
     \bigl(-if(t) + ig'(t) \bigr)D_Y\phi ^+_\lambda .  $$
In this case the involution
  $$ f(t)\phi ^+_\lambda + g(t)D_Y\phi ^+_\lambda \longmapsto \sqrt\lambda
     g(t)\phi ^+_\lambda  + \frac{f(t)}{\sqrt\lambda }D_Y\phi ^+_\lambda
      $$
anticommutes with~$D_C$ and preserves the boundary conditions~\thetag{1.2},
which reduce to the equations
  $$ \aligned
      g(1) + \frac{f(1)}{\sqrt\lambda } &=0\\
      f(-1) + \sqrt\lambda g(-1) &= 0\endaligned \tag{4.11} $$
Therefore, the part of the spectrum of~$D_C$ coming from spinor
fields~\thetag{4.9} with~$\lambda >0$ is symmetric about the origin, so does
not contribute to the $\eta $-invariant.  An easy computation shows that
$\Ker D_C$ contains no nonzero spinor fields which are sums of fields of the
form~\thetag{4.10} subject to the boundary constraint~\thetag{4.11}.  So there
is no contribution to the $\xi $-invariant.

We are left to consider spinor fields
  $$ \psi =f(t)\phi ^+ + g(t)\phi ^-,\qquad \phi \in K^+,\quad \phi ^-\in
     K^-,  $$
subject to the boundary condition
  $$ \pmatrix f(-1)\phi ^+\\g(1)\phi ^-  \endpmatrix + T\pmatrix f(1)\phi
     ^+\\g(-1)\phi ^-  \endpmatrix =0. \tag{4.12} $$
Now
  $$ D_C\psi  = -if'(t)\php + ig'(t)\phm,  $$
and it is straightforward to see that $D_C\psi =\mu \psi $ subject
to~\thetag{4.12} if and only if
  $$ \psi =e^{i\mu t}\php + e^{-i\mu t}\phm  $$
with
  $$ T\pmatrix \php \\ \phm  \endpmatrix = -e^{-2i\mu }\pmatrix \php \\ \phm
     \endpmatrix.  $$
So each eigenvalue~$\nu $ of~$T$ contributes a set of the form~$\mu +\pi \ZZ$
to the spectrum of~$D_C$, where~$0\le\mu <\pi $ satisfies $-e^{-2i\mu } =\nu
$.  A standard computation (e.g.~\cite{APS}) shows that the $\eta $-invariant
of the set~$\mu +\pi \ZZ$ is~$1-\dfrac{2\mu }{\pi }$ if~$\mu \not= 0$.  Thus
if~$\mu \not= 0$ the $\xi $-invariant is~$\dfrac 12 -\dfrac \mu \pi $, and
exponentiating we obtain $e^{\tpi(\frac 12 - \frac \mu \pi )} = -e^{-2i\mu }
= \nu .$ This is the correct value of the exponentiated $\xi $-invariant
for~$\mu =0$ as well.  Combining the contribution from all of the eigenvalues
we obtain~\thetag{4.8}.

Now for $a>0$ the boundary condition is a unitary map
  $$ T\:K^+_Y(a) \oplus K^-_Y(a)\longrightarrow K^+_Y(a) \oplus K^-_Y(a).
     \tag{4.13} $$
If $T=T_0$~has the form $T_0=T'\oplus \dfrac{D}{\sqrt{D^2}}$
for~$D=D_{\partial C}(\epsilon ,a)$ and some isometry $T'\:K^+_Y(\epsilon)
\oplus K^-_Y(\epsilon)\to K^+_Y(\epsilon) \oplus K^-_Y(\epsilon )$, then the
desired result follows from the previous argument and~\thetag{1.6}.  (Recall
that \thetag{1.6}~is a triviality.)  Another isometry~$T$~\thetag{4.13} is
connected to~$T_0$ via a path of isometries~$T_t$, and by \theprotag{3.3}
{Theorem} and~\thetag{3.2} we have
  $$ \frac{1}{\tau _C(a,T_t)}\frac{d\tau _C(a,T_t)}{dt} = \frac{1}{\det
     T_t}\frac{d(\det T_t)}{dt}.  $$
It follows that $\tau _C(a,T_t)=\det T_t$ as desired.
        \enddemo

        \demo{Proof of \theprotag{4.5} {Proposition}}
 Following Bunke~\cite{B1} we will first construct an isometry
  $$ U\:H_{X\cut}(a',T';a,T)\longrightarrow H_X(a',T')\oplus
     H_C(a,\tilde{T}), \tag{4.14} $$
where the notation means the subspace of spinor fields which satisfy the
appropriate boundary condition~\thetag{1.2}.  Note the appearance of
  $$ \tilde{T} = \pmatrix 1&\\&-1  \endpmatrix T\pmatrix -1&\\&1
     \endpmatrix. \tag{4.15} $$
We then compute
  $$ Q=U\inv (D_X\oplus D_C)U - D_{X\cut}, \tag{4.16} $$
which turns out to be a bundle endomorphism supported on the disjoint union
of two cylinders.  It follows that
  $$ \frac{d}{du}e^{\tpi\xi (D_{X\cut} + uQ)} \tag{4.17} $$
may be computed locally, and we use a symmetry argument to prove that it
vanishes.  Equating the values at~$u=0$ and~$u=1$ we see that
  $$ \tau _{X\cut}(a',T';a,T) = \tau _X(a',T')\tau _C(a,\tilde{T}),
     \tag{4.18} $$
which reduces to~\thetag{4.6} using~\thetag{4.8}.


To begin let $f_L,f_R\:\moo\to\zo$ be smooth cutoff functions which satisfy
(Figure~2)
  $$ \gathered
      f_L\bigl([-1,-1/2] \bigr) = f_R\bigl([1/2,1] \bigr) = 1\\
      f_L\bigl([1/2,1] \bigr) = f_R\bigl([-1,-1/2] \bigr) = 0\\
      f_L^2 + f_R^2 = 1\\
      f_L(-x) = f_R(x).\endgathered\tag{4.19} $$
The functions~$f_L,f_R$ lift to functions on $C=\moo\times Y$.

As in Figure~3 we choose isometric embeddings $C\hookrightarrow X\cut$ near
the boundary pieces~$Y$ and~$-Y$.  Denote the image cylinders by~$C_1$
and~$C_2$ respectively.  Similarly, we choose an isometric embedding
$C\hookrightarrow X$ with image~$C_3$ so that we obtain~$X\cut$ from~$X$ by
cutting along $\{0\}\times Y\subset C_3$.  If we cut~$X\cut$
along~$\{0\}\times Y\subset C_1$ and~$\{0\}\times Y\subset C_2$ then two
extra pieces fall out, and they reassemble to form an extra cylinder~$C_4$.
Define~$U$ as follows.  Let $\psi $~be a spinor field on~$X\cut$.  Let~$U$
map its restriction to the complement of~$C_1\sqcup C_2$ unchanged to the
complement of~$C_3$ in~$X$.  Then let $\psi _1,\psi _2$~be the restrictions
of~$\psi $ to~$C_1,C_2$, and define
  $$ U\:\pmatrix \psi _1\\\psi _2 \endpmatrix\longmapsto \pmatrix
     f_L&f_R\\-f_R&f_L \endpmatrix\pmatrix \psi _1\\\psi _2 \endpmatrix.
     \tag{4.20} $$
The right hand side of~\thetag{4.20} is an element of~$H_{C_3}\oplus
H_{C_4}$, and it patches to~$\psi $ on~$X-C_3$ to give a smooth spinor field
on~$X\sqcup C_4$.  Note the change in the boundary values on~$C_4$, as
indicated in~\thetag{4.14} and~\thetag{4.15}.  It is easy to check that
$U$~is unitary.


Next we compute~$Q$, which is defined in~\thetag{4.16}.  Since $U$~is the
identity on the complement of~$C_1\sqcup C_2$, the operator~$Q$ has support
on~$C_1\sqcup C_2$.  An easy computation yields
  $$ Q\:\pmatrix \psi _1\\\psi _2 \endpmatrix\longmapsto \pmatrix 0&\theta
     \\-\theta &0 \endpmatrix\pmatrix \psi _1\\\psi _2 \endpmatrix,
      $$
where the 1-form
  $$ \theta =f_Ldf_R - f_Rdf_L  $$
acts by Clifford multiplication.  Notice that $\theta $~is supported in the
interior of~$C_1\sqcup C_2$.

Consider the map
  $$ I\:\pmatrix \psi _1\\\psi _2  \endpmatrix \longmapsto \pmatrix dx\cdot
     \psi _2(-x) \\dx\cdot \psi _1(-x)  \endpmatrix,  $$
where `$\cdot $'~denotes Clifford multiplication.  This is the map on spinor
fields induced by the orientation preserving diffeomorphism $\langle x_1,x_2
\rangle\mapsto\langle -x_2,x_1  \rangle$ of~$C_1\sqcup C_2$.  We only
apply~$I$ on the domain of~$Q$, so we need only consider~$\langle \psi
_1,\psi _2  \rangle$ with support in the interior of~$C_1\sqcup C_2$.  It is
easy to verify
  $$ \aligned
      I^2&=-1\\
      ID&=-DI\\
      IQ&=-QI,     \endaligned \tag{4.21} $$
where $D=D_C$ is the Dirac operator on~$C$.  For the second equation, note
that any orientation-reversing isometry anticommutes with the Dirac operator.
For the third, note that
  $$ \theta (-x) = \theta (x)  $$
from equations~\thetag{4.19}.

Let $\xi _u$ denote the $\xi $-invariant of~$D_u=D_{X\cut} + uQ$.  As in
\theprotag{3.10} {Lemma} its variation is computed by the formula
  $$ \frac{d\xi _u}{du} = \frac{-1}{\sqrt\pi }\lim\limits_{t\to0}t^{1/2}
     \Tr_{X\cut}(Qe^{-tD_u^2}).  $$
Now the right hand side is the integral over~$X\cut$ of a locally computed
quantity, and since $Q$~has support in~$C_1\sqcup C_2$ the integral may be
computed there.  But from~\thetag{4.21} we have
  $$ \split
      \Tr(Q\etD) &= -\Tr(I^2Q\etD)\\
      &=\hphantom{-}\Tr(IQI\etD)\\
      &=\hphantom{-}\Tr(IQ\etD I)\\
      &=\hphantom{-}\Tr(I^2Q\etD)\\
      &=-\Tr(Q\etD).     \endsplit  $$
This proves that \thetag{4.17}~vanishes, from which~\thetag{4.18} and then
\thetag{4.6}~follow.
        \enddemo

As a corollary of \theprotag{4.5} {Proposition} we derive~\thetag{1.5},
which is a generalization of~\cite{LW,Theorem~3.1}.

        \proclaim{\protag{4.22} {Corollary}}
 Let $X$~be a compact odd dimensional spin manifold with boundary.  Choose
a positive number $a\notin \spec(D\db^2)$ and isometries
$T_1,T_2\:K\db^+(a)\to K\db^-(a)$.  Then
  $$ \tau _X(a,T\mstrut _2) = \det(T_1\inv T\mstrut _2) \,\tau _X(a,T\mstrut
     _1). \tag{4.23} $$
        \endproclaim

        \demo{Proof}
 Let $C=\moo\times \bX\hookrightarrow X$ be an isometric embedding mapping
$\{1\}\times \bX$ onto~$\bX$, and let $Y$~be the image of $\{0\}\times \bX$.
Cutting along~$Y$ we obtain~$X\cut$ which is (spin) isometric to~$X\sqcup C$.
Consider the boundary conditions defined by~$T_2$ on~$\bX$.  On~$Y\sqcup -Y$
we use the boundary conditions
  $$ T=\pmatrix 0&T_1\inv \\T\mstrut _2&0  \endpmatrix.  $$
Note that
  $$ \det T = (-1)^{\dim K\db^+(a)}\det(T_1\inv T\mstrut _2).  $$
The induced boundary conditions on~$C$ are
  $$ \tilde{T} = \pmatrix 0&T_1\inv \\T\mstrut _1&0  \endpmatrix,
     $$
and
  $$ \det \tilde{T} = (-1)^{\dim K\db^+(a)}.  $$
Now \thetag{4.6}~and \thetag{4.8}~imply the desired result~\thetag{4.23}.
        \enddemo

\newpage
\head
\S{5} Adiabatic limits and holonomy
\endhead
\comment
lasteqno 5@ 12
\endcomment

In this section we reprove the main result in~\cite{BF2} which computes the
holonomy of the natural connection~$\nabla $ on the (inverse) determinant
line bundle as the {\it adiabatic limit\/} of exponentiated $\xi $-invariants
(on a closed manifold).  Our proof here uses the curvature
formula\footnote{In fact, it suffices to consider the case where the base~$Z$
is a circle, and then the curvature obviously vanishes.  So the curvature
formula is not really needed.} proved in~\cite{BF1}, ~\cite{BF2}, the
variation formula~\thetag{1.10}, and the gluing law~\thetag{2.21}.  We define
a new connection~$\nabla '$ by specifying its {\it parallel transport\/} as
the adiabatic limit of exponentiated $\xi $-invariants, now defined on
manifolds with boundary.  We then show that~$\nabla '=\nabla $.

Let $\pi \:Y\to Z$ be a spin map whose typical fiber is a closed even
dimensional manifold, and let $L\to Z$ denote the inverse determinant line
bundle.  According to~\cite{BF1} it comes equipped with a (Quillen) metric
and a natural unitary connection~$\nabla $.  The curvature\footnote{Since we
use the {\it inverse\/} determinant line bundle the sign in~\thetag{5.1}
differs from that in~\cite{BF2}.} of~$\nabla $ is~\cite{BF2,Theorem~1.21}
  $$ \Omega ^L = -\tpi \comp{\int_{Y/Z}\Ahat(\Omega ^{Y/Z})}{2}. \tag{5.1} $$

We now define~$\nabla '$.  Let $\path$~denote the space of smooth
parametrized paths $\gamma \:\zo\to Z$ with $\gamma \res{[0,0.1]}$ and
$\gamma \res{[0.9,1]}$ constant.  For~$\gamma \in \path$ let $Y_\gamma
=\gamma ^*Y$ denote the pullback of $\pi \:Y\to Z$ via~$\gamma $; then $\pi
_\gamma \:Y_\gamma \to\zo$ is a spin map.  Let $g_{\zo}$~denote an arbitrary
metric on~$\zo$ and $g_{Y/Z}$~the metric on the relative tangent
bundle~$T(Y/Z)$.  Define a family of metrics on~$Y_\gamma $ by the formula
  $$ g_\epsilon = \frac{g_{\zo}}{\epsilon ^2}\oplus g_{Y/Z},\qquad \epsilon
     \not= 0. \tag{5.2} $$
The metric~$g_\epsilon $ on~$Y_\gamma $ is determined by requiring that $\pi
_\gamma $~be a Riemannian submersion.  Physicists term
`$\lim\limits_{\epsilon \to 0}$' the {\it adiabatic limit\/}.  The spin
structure on~$T(Y_\gamma /Z)$ induces one on~$TY_\gamma $ since
  $$ TY_\gamma \cong \pi _\gamma ^*T(\zo)\oplus T(Y/Z) \tag{5.3} $$
and the latter factor is trivial.  Now the exponentiated $\xi $-invariant is
a map
  $$ \tau _{Y_\gamma }(\epsilon )\:L_{\gamma (0)}\longrightarrow L_{\gamma
     (1)}. $$
Here we use the isomorphisms~\thetag{2.17} and~\thetag{2.18}.

        \proclaim{\protag{5.4} {Lemma}}
 The adiabatic limit $\tau _\gamma =\lim\limits_{\epsilon \to0}\tau _{Y_\gamma
}(\epsilon )$ exists and is independent of the choice of \break
metric~$g_{\zo}$.
        \endproclaim

        \demo{Proof}
 As a preliminary we state without proof a simple result about the Riemannian
geometry of adiabatic limits.  Let $\nabla ^{Y_\gamma }(\epsilon )$~denote
the Levi-Civita connection on~$\Yg$ with the metric~\thetag{5.2} and
$\curv{\Yg}(\epsilon )$~its curvature.  Then $\lim\limits_{\epsilon
\to0}\nabla
^{\Yg}(\epsilon )$~exists and is torsionfree.  Furthermore, the curvature of
this limiting connection is the limit of the curvatures of~$\nabla
^{\Yg}(\epsilon )$ and has the form
  $$ \lim\limits_{\epsilon \to0}\curv{\Yg}(\epsilon ) = \pmatrix 0&0\\
     *&\curv{\Yg/\zo} \endpmatrix \tag{5.5} $$
relative to the decomposition~\thetag{5.3}.  We will apply this result in
families, where it also holds.

Consider the spin map $p\:\Yg\times (\RR-\{0\})\to\RR-\{0\}$, where the
metric on the fiber at~$\epsilon $ is~\thetag{5.2}.  According to
\theprotag{1.9} {Theorem} we have
  $$ \frac{d}{d\epsilon }\tau _{\Yg}(\epsilon ) =
     \tpi\comp{\int_{p}\Ahat(\curv p)}{1}.  $$
Now \thetag{5.5}~immediately implies that the component of the integrand in
the $\zo$~direction approaches zero as~$\epsilon \to0$.  In other words, if
$t$~is the coordinate in the $\zo$~direction, then any term in the integrand
involving~$dt$ approaches zero as~$\epsilon \to0$.  Hence
$\lim\limits_{\epsilon \to0}\dfrac{d}{d\epsilon }\tau _{\Yg}(\epsilon )=0$ and
so $\lim\limits_{\epsilon \to0}\tau _{\Yg}(\epsilon )$ exists.

A similar argument proves that $\tau _\gamma $ is independent of~$g_{\zo}$.
Let $\Cal{M}$~denote the space of metrics on~$\zo$ and consider the spin map
  $$ \Yg\times (\RR-\{0\})\times \Cal{M}\longrightarrow (\RR-\{0\})\times
     \Cal{M},  $$
where the metric on the fiber over~$\langle \epsilon ,g_{\zo}  \rangle$
is~\thetag{5.2}.  As in the previous argument we see that the differential
of~$\tau _{\Yg}(\epsilon ,g_{\zo})$ with respect to~$g_{\zo}$ vanishes
as~$\epsilon \to0$.  The desired conclusion follows immediately.
        \enddemo

An immediate corollary is that $\tau _\gamma $~is invariant under
reparametrization of paths.  Also, if $\gamma _1,\gamma _2\in \path$ with
$\gamma _1(1) = \gamma _2(0)$, and $\gamma _2\circ \gamma _1$~denotes the
composed path, then $\tau _{\gamma _2\circ \gamma _1} = \tau _{\gamma
_2}\circ \tau _{\gamma _1}$.  This follows from the gluing law
(\theprotag{2.20} {Theorem}).  Now a general theorem~\cite{F2,Appendix~B}
applies to construct a connection~$\nabla '$ on~$L$ whose parallel transport
is~$\tau $.

Now we compute the holonomy of~$\nabla '$.  Let $\gb\:\cir\to Z$ be a loop
in~$Z$ and $Y_{\gb}\to\cir$ the corresponding fibered manifold.
Realize~$\gb$ as the gluing of a path $\gamma \:\zo\to Z$; then $Y_{\gb}$~is
obtained by identifying the ends of $Y_\gamma \to\zo$.  This identification
induces the spin structure on~$Y_{\gb}$ obtained by lifting the {\it
nonbounding\/} spin structure on~$\cir$.  The gluing law \theprotag{2.20}
{Theorem} implies (compare~\thetag{2.24})
  $$ \split
      \lim_{\epsilon \to0}\tau _{Y_{\gb}}(\epsilon ) &=
     \Tr_Y\bigl(\lim_{\epsilon \to0} \tau _{Y_\gamma }(\epsilon ) \bigr)\\
      &= \Tr_Y(\text{parallel transport along $\gamma$} )\\
      &= (-1)^{\index D_Y}\cdot (\text{holonomy around $\gb$} ).
     \endsplit \tag{5.6} $$
If $L=L_{\gamma (0)}$ then the parallel transport is an element of $L\gtimes
L^*$.  The sign comes since the composition $L\gtimes L^* \to L^*\gtimes L\to
\CC$ is $(-1)^{|L|}=(-1)^{\index D_Y}$ times the usual contraction.  Let
$Y_{\gb}'$ denote~$Y\mstrut _{\gb}$ with spin structure induced by lifting
the {\it bounding\/} spin structure on~$\cir$.  If we substitute~$Y'_{\gb}$
for~$Y\mstrut _{\gb}$ in~\thetag{5.6}, then the resulting equation has no
factor~$\sind$.  This follows as in~\thetag{2.25}.
(Compare~\cite{F1,Theorem~1.31}.)

Our main result in this section is the following.

        \proclaim{\protag{5.7} {Proposition}}
 $\nabla '=\nabla $.
        \endproclaim

To prove \theprotag{5.7} {Proposition} we compare the covariant derivative
of their parallel transports using the following general lemma.

        \proclaim{\protag{5.8} {Lemma}}
 Let $L\to Z$ be an arbitrary line bundle with connection~$\nabla $ and
curvature~$\curv L$.  Denote the parallel transport of~$\nabla $ along a
path~$\gamma $ by~$\rho _\gamma $.  Then
  $$ \nabla \rho = - \bigl(\int_{p_2}\ev^*\curv L \bigr)\cdot \rho ,
     \tag{5.9} $$
where $\ev$ and $p_2$~are the maps
  $$ \CD
     [0,1]\times \path @>\ev>> Z\\
     @Vp_2VV \\
     \path\endCD  $$
        \endproclaim

\flushpar
 To interpret~\thetag{5.9} view $\rho $ as a section of the line bundle
$\bigl(\ev_0^*(L) \bigr)^*\otimes \bigl(\ev_1^*(L) \bigr)\to\path$ with its
connection induced from~$\nabla $.  Here $\ev_t(\gamma ) = \gamma (t)$.  The
proof is elementary.

        \proclaim{\protag{5.10} {Corollary}}
 If $\nabla ,\nabla '$ are connections on $L\to Z$ with parallel
transports~$\rho ,\tau $, and if $\dfrac{\nabla \rho}{\rho } =\dfrac{\nabla
\tau }{\tau }$, then~$\nabla '=\nabla $.
        \endproclaim

\flushpar
 For if $\nabla '=\nabla +\alpha $ for a 1-form $\alpha $ on~$Z$, then
  $$ \frac{\nabla \tau }{\tau }- \frac{\nabla \rho }{\rho }= -
     \bigl(d\int_{p_2}\ev^*\alpha \bigr) , $$
and if $\alpha \not= 0$ then the right hand side is nonzero.

We now verify the hypotheses of \theprotag{5.10} {Corollary} for the natural
connection~$\nabla $ and the new connection~$\nabla '$ on the inverse
determinant line bundle.  We use the diagram
  $$ \CD
     ev^*Y @>>> Y\\
     @V\pi 'VV @VV\pi V\\
     [0,1]\times \path @>\ev>> Z\\
     @Vp_2VV \\
     \path\endCD   $$
We compute $\nabla \tau $ using the variation formula~\thetag{1.10}.  Namely,
$\tau _\gamma $~is the adiabatic limit of~$\tau _{\Yg}$, and $\Yg$~is the
fiber $(p_2\circ \pi ')\inv (\gamma )$.  So by the variation formula
  $$ \split
     \nabla \tau &= \tpi\comp{\int_{p_2\circ \pi '}
     \alim\Ahat(\curv{p_2\circ \pi '})}{1}\cdot \tau  \\
     &=\tpi\int_{p_2 } \comp{\int_{\pi '}\Ahat(\curv{\pi '})}{2}\cdot \tau\\
     &=\int_{p_2 } \ev^*\comp{\tpi\int_{\pi }\Ahat(\curv{\pi})}{2}\cdot
     \tau,\endsplit  $$
where we use~\thetag{5.5} to pass from the first equation to the second.  (Of
course, `$\alim$' is the adiabatic limit.)  But by~\thetag{5.9} and the
curvature formula~\thetag{5.1} this latter expression is the covariant
derivative of the parallel transport of~$\nabla $.  This concludes the proof
of \theprotag{5.7} {Proposition}.

Therefore, \thetag{5.6}~also computes the holonomy of the canonical
connection~$\nabla $ on the inverse determinant line bundle as the adiabatic
limit of exponentiated $\xi $-invariants.  This is exactly\footnote{Again,
since we use the {\it inverse\/} determinant line bundle the sign
in~\thetag{5.12} differs from~\cite{BF2}.} the content of Theorem~3.16
in~\cite{BF2}:

        \proclaim{\protag{5.11} {Corollary}}
 Let $\gb\:\cir\to Z$ be a loop and $Y_{\gb}\to \cirnonbdd$ the corresponding
fibered manifold.  Then the holonomy around~$\gb$ of the natural
connection~$\nabla $ on the inverse determinant line bundle~$L\to Z$ is
  $$ \sind\alim\left( e^{\tpi\xi _{Y_{\gb}}} \right).  \tag{5.12} $$
        \endproclaim

\newpage
\head
\S{6} Remarks on the Families Index Theorem
\endhead
\comment
lasteqno 6@ 12
\endcomment

Let $\pi : X \to Z$ be a spin map whose typical fiber is a compact even
dimensional manifold with boundary. When $\ker D_{\partial X_z}$ has constant
rank there is a well-defined index bundle $\operatorname{Ind} D_{X/Z} \in
K^0(Z)$.  The families index theorem of Bismut-Cheeger states that its Chern
character $\ch(\operatorname{Ind} D_{X/Z})$ is represented in de Rham
cohomology by (cf.~\cite{BC2},~\cite{BC3}, ~\cite{D3}, a more general version
has been proved in~\cite{MP})
  $$  \int_{X/Z} \hat{A} (\Omega^{X/Z}) - \tilde{\eta},  \tag{6.1} $$
where $\tilde{\eta}$ is a differential form on the base $Z$, defined as
follows.

Consider a spin map $\pi \:Y\to Z$ whose typical fiber is a closed manifold.
(Our application takes~$Y=\bX$.)  The associated Bismut superconnection $A_t$
is
  $$ A_t =\tilde{\nabla} + t^{1/2}D_{Y/Z} - \frac{c(T)}{4t^{1/2}},
      $$
where $c(T) = \sum_{\alpha \leq \beta} dz^{\alpha}\,dz^{\beta} T(f_{\alpha}
,f_{\beta})$ with $T$ the curvature form of the fibration, $f_{\alpha}$ a
local orthonormal basis on $Z$, and $dz^{\alpha}$ the 1-form dual to
$f_{\alpha}$.  The asymptotics of heat kernels associated to the Bismut
superconnection exhibit some remarkable cancellations. The first one is
expressed in the local index theorem for families~\cite{Bis}, \cite{BF2}.
More essential to our discussion are two other cancellation
results~\cite{BC1}:
  $$ \align
      tr_s[(D_{Y/Z} + \frac{c(T)}{4t}) e^{-A_t^2}] &= O(t^{1/2})\ \text{as $t
     \to 0$}, \qquad \text{if $\dim Y/Z$ is even;} \tag{6.2} \\
      tr^{\text{even}}[(D_{Y/Z} + \frac{c(T)}{4t}) e^{-A_t^2}] &= O(t^{1/2})
     \ \text{as $t \to 0$}, \qquad \text{if $\dim Y/Z$ is odd.}\tag{6.3}
     \endalign $$ 
where $tr^{\text{even}}$ indicates the even form part of $tr$.  When $\ker
D_{Y/Z}$ has constant rank, the expressions on the left hand
sides of~\thetag{6.2}, ~\thetag{6.3} are also well behaved for the large
time. In fact, it is shown in~\cite{BGV} (in a more general setting) that
  $$ \align
      tr_s[(D_{Y/Z} + \frac{c(T)}{4t}) e^{-A_t^2}] &= O(t^{-1}) \ \text{as
     $t\to \infty $}, \qquad \text{if $\dim Y/Z$ is even;} \tag{6.4} \\
      tr^{\text{even}}[(D_{Y/Z} + \frac{c(T)}{4t}) e^{-A_{t}^{2}}] &=
     O(t^{-1}) \ \text{as $t \to \infty $}, \qquad \text{if $\dim Y/Z$ is
     odd.} \tag{6.5} \endalign $$

By virtue of~\thetag{6.2}--\thetag{6.5} we now define a differential form on
$Z$, the $\hat{\eta}$ form:
  $$ \hat{\eta} =\cases \dfrac{1}{\sqrt{\pi}} {\displaystyle
     \int}_{0}^{\infty} tr_{s}[(D_{Y/Z} + \dfrac{c(T)}{4t})
     e^{-A_{t}^{2}}] \,\dfrac{dt}{2t^{1/2}},&\text{if $\dim Y/Z$ is
     even} ;\\\dfrac{1}{\sqrt{\pi}} {\displaystyle \int}_{0}^{\infty}
     tr^{even}[(D_{Y/Z} + \dfrac{c(T)}{4t}) e^{-A_{t}^{2}}]
     \,\dfrac{dt}{2t^{1/2}} ,&\text{if $\dim Y/Z$ is odd}
     .\endcases $$
For example, the first integral is convergent at $0$ because of~\thetag{6.2},
and convergent at $\infty$ because of~\thetag{6.4}.  We normalize
$\hat{\eta}$ by defining
  $$ \tilde{\eta} =\cases {\displaystyle \sum \frac{1}{(2\pi i)^j}
     \comp{\hat{\eta}}{2j-1}},&\text{if $\dim Y/Z$ is even}
     ;\\{\displaystyle \sum \frac{1}{(2\pi i)^j} \comp{\hat{\eta}}{2j}}
     ,&\text{if $\dim Y/Z$ is odd} .\endcases $$
Here we decompose the odd (respectively even) form $\hat{\eta}$ into its
homogeneous components $\comp{\hat{\eta}}{2j-1}$ (respectively
$\comp{\hat{\eta}}{2j}$).  The $\tilde{\eta }$~form satisfies a transgression
formula.  If $\dim Y/Z$~is odd, then~\cite{BC2}, \cite{D2}
  $$ d \tilde{\eta}=-\int_{Y/Z} \hat{A} (\Omega^{Y/Z}). \tag{6.6} $$
If $\dim Y/Z$~is even and $\ker D_Y$~has constant rank, then~\cite{D2}
  $$ d \tilde{\eta}= \ch(\operatorname{Ind} D_{Y/Z}) -\int_{\partial
     X/Z} \hat{A} (\Omega^{Y/Z}). \tag{6.7} $$

Return now to a spin map $\pi \:X\to Z$ whose typical fiber is a compact
manifold with boundary.  If $\dim X/Z$~is even, which is the case considered
by Bismut-Cheeger, then \thetag{6.6}~immediately implies that the
differential form~\thetag{6.1} is closed.  We are interested in the case
where $\dim X/Z$~is odd, and then \thetag{6.7}~implies that unless
$D_{\bX/Z}$ is invertible, the differential form~\thetag{6.1} is {\it not\/}
closed. Thus in the odd dimensional case one expects a correction term in the
Bismut-Cheeger index formula from $\ker D_{\partial X/Z}$.

\theprotag{3.3} {Theorem} suggests what the correction term should be,
assuming that $\ker D_{\bX/Z}$~has constant rank. To define the odd index
bundle we need self-adjoint operators. In our case this amounts to a choice
of a (smooth) family of isometries
  $$ T\: \ker D^+_{\partial X/Z} \longrightarrow \ker D^-_{\partial X/Z}. $$
The resulting family of self-adjoint operators $D_{X/Z}(T)$ gives rise to a
well-defined index bundle $\operatorname{Ind} D_{X/Z}(T) \in K^1(Z)$.  On the
other hand, $\ch(\operatorname{Ind} D_{\partial X/Z})=\operatorname{Tr}_s
(e^{-(\nabla^a)^2})$, where $a$ is chosen to be smaller than the smallest
eigenvalue of $D_{\partial X/Z}$. Consider the superconnection $\nabla^a +
\sqrt{t} V$ on $\ker D_{\partial X/Z}$, with $V$ the symmetric endomorphism
  $$ V= \pmatrix 0 & T^* \\ T & 0 \endpmatrix. $$
One has the following transgression formula
  $$ \frac{d}{dt} \operatorname{Tr}_s(e^{-(\nabla^a + \sqrt{t} V)^2}) = - d [
     \frac{1}{2\sqrt{t}}\operatorname{Tr}_s(Ve^{-(\nabla^a + \sqrt{t} V)^2})
     ], $$
which, by the invertibility of $V$, yields
  $$ d\tilde{\eta}_T = \ch(\operatorname{Ind} D_{\partial X/Z}),  $$
with  $\tilde{\eta}_T$ defined by
  $$ \tilde{\eta}_T=\int_0^{\infty}
     \frac{1}{2\sqrt{t}}\operatorname{Tr}_s(Ve^{-(\nabla^a + \sqrt{t} V)^2})
     \,dt.  $$

        \proclaim{\protag{6.8} {Conjecture}}
 The (odd) Chern character of $\operatorname{Ind} D_{X/Z}(T)$ is represented
in the de Rham cohomology by
  $$ \int_{X/Z} \hat{A} (\Omega^{X/Z}) - \tilde{\eta} - \tilde{\eta}_T $$
        \endproclaim

We have the following evidence for this conjecture.

        \proclaim{\protag{6.9} {Theorem}}
 The degree one component of the odd Chern character of the index bundle
\break
$\ch_1(\operatorname{Ind} D_{X/Z}(T)) \in H^1(Z)$ is represented by
  $$ \comp{\int_{X/Z} \hat{A} (\Omega^{X/Z}) - \tilde{\eta} -
     \tilde{\eta}_T}{1}.  $$
        \endproclaim

        \demo{Proof}
 By the Duhamel principle
  $$ \comp{\operatorname{Tr}_s(Ve^{-(\nabla^a + \sqrt{t} V)^2})}{1}=
     -\sqrt{t} \operatorname{Tr}_s(V (\nabla^a V)e^{-t V^2}). $$
Therefore,
  $$ \split \comp{ \tilde{\eta}_T}{1} & = -\int_0^{\infty}
     \frac{1}{2}\operatorname{Tr}_s(V(\nabla^a V)e^{-t V^2}) dt \\& =
     -\frac{1}{2}\operatorname{Tr}_s(V^{-1}\nabla^a V) \\&=
     -\operatorname{Tr}(T^{-1}\nabla^a T). \endsplit \tag{6.10}$$
Similarly, we have
  $$ \comp{\tilde{\eta}}{1} =-\frac{1}{2} \int_0^\infty \operatorname{Tr}_s
     (D_{\partial X/Z} \tilde{\nabla} D_{\partial X/Z} e^{-t D_{\partial
     X/Z}^2}) \,dt. \tag{6.11} $$

On the other hand, the degree one component of $\ch(\operatorname{Ind}
D_{X/Z}(T))$ is given by $d\xi_X(a, T)$, which, according to
Theorem~\thetag{3.3}, gives
  $$ \ch_1(\operatorname{Ind} D_{X/Z}(T)) = \comp{\int_{X/Z} \hat{A}
     (\Omega^{X/Z})}{1} + \frac{1}{2\pi i} \,u^{-1} \nabla u.  $$
{}From~\thetag{3.24}, ~\thetag{3.25}, ~\thetag{3.26} and our choice of $a$ we
have
  $$ \split u^{-1} \nabla u= (\Det T)^{-1}\, \nabla^a (\Det T) & + \LIM_{t
     \to 0} \operatorname{Tr} \bigl((D^+)^{-1} \tilde{\nabla} D^+ e^{-t
     D^2}\bigr) \\ &\qquad - \frac{1}{2}\LIM_{t \to 0} \operatorname{Tr}
     \bigl((D^2)^{-1} \tilde{\nabla}(D^2) e^{-t D^2}\bigr), \endsplit
     \tag{6.12} $$
and the first term in~\thetag{6.12} is exactly $-\comp{ \tilde{\eta}_T}{1}$
by~\thetag{6.10}. For the remaining terms we note from~\thetag{6.11}
and~\thetag{3.23}
  $$ \split [\tilde{\eta}]_{(1)} &= -\frac{1}{2} \LIM_{t \to 0}
     \operatorname{Tr}_s \bigl[ D^{-1} \tilde{\nabla} D e^{-t D^2}\bigr] \\
     & = -\frac{1}{2} \LIM_{t \to 0} \operatorname{Tr} \bigl[
     (D^+)^{-1} \tilde{\nabla} D^+ e^{-t D^2}\bigr] + \frac{1}{2} \LIM_{t \to
     0} \operatorname{Tr} \bigl[ (D^-)^{-1} \tilde{\nabla} D^- e^{-t
     D^2}\bigr] \\ & = -\LIM_{t \to 0} \operatorname{Tr} \bigl[
     (D^+)^{-1} \tilde{\nabla} D^+ e^{-t D^2}\bigr] + \frac{1}{2}\LIM_{t \to
     0}\operatorname{Tr} \bigl[(D^2)^{-1} \tilde{\nabla}(D^2) e^{-t
     D^2}\bigr]. \endsplit$$
This finishes the proof.

  \enddemo

\newpage
\head
\S{A} Appendix: Generalized APS Boundary Conditions
\endhead
\comment
lasteqno A@ 17
\endcomment

In this appendix we discuss the analytical aspects of the generalized APS
boundary conditions. For simplicity of notation we restrict ourself to the
case of Dirac operators, although our discussion extends easily to the more
general situation of Dirac-type operators.

Let $X$ be an odd dimensional compact oriented spin manifold with smooth
boundary $\partial X = Y$. We shall always assume that the Riemannian metric
on $X$ is a product near the boundary. Let
  $$  D: \ C^\infty(X, S) \rightarrow C^\infty(X, S) $$
be the formally self-adjoint Dirac operator acting on the spinor bundle $S
\to X$. Then in a collar neighborhood $[0,1) \times \bX $ of the boundary,
$D$~takes the form
  $$  D = J( \partial_u + D_{\partial X}), $$
where $J = c(du)$ and
  $$  D_{\partial X}: \ C^\infty(\bX , S|_{\bX} ) \rightarrow C^\infty({\bX} ,
S|_{\bX} ) $$
is the self-adjoint Dirac operator on $\bX $ under the identification $S\res
{\bX}
\cong S(\bX )$.

As an unbounded operator in $L^2(X, S)$ with domain $C_0^\infty(X, S)$, $D$
is symmetric. (In other words, $D$ is formally self-adjoint). To obtain
self-adjoint extensions of $D$, one has to impose boundary conditions. For
our purpose, we would like to restrict our attention to boundary conditions
of elliptic type. Appropriate boundary conditions that are of elliptic type
are considered by Atiyah-Patodi-Singer~\cite{APS}.  Namely if we denote by
$\Pi_+$ the orthogonal projection of $L^2(\bX , S|_{\bX} )$ onto the subspace
spanned
by the eigensections of $D_{\partial X}$ with nonnegative eigenvalues, then
$D_+ = D$ with domain
  $$ \text{dom} (D_+) = \{ \varphi \in H^1 (X, S) \bigm| \Pi_+ (\varphi
     |_{\bX} ) = 0 \} $$
is an elliptic boundary value problem (in the generalized sense,
see~\cite{APS}, \cite{Se}). $D_+$ is a closed symmetric extension of $D$,
although, in general, $D_+$ is not self-adjoint. However, one can obtain
elliptic self-adjoint boundary value problems by considering further
self-adjoint extensions of $D_+$.

More generally, let $a \not\in \text{spec} D_{\partial X}^2$ be a positive
number and $\Pi_{-a}$ (respectively $\Pi_a$) denote the orthogonal projection
of $L^2(\bX , S|_{\bX} )$ onto the subspace spanned by eigensections of
$D_{\partial
X}$ with eigenvalues $> -\sqrt{a}$ (respectively $> \sqrt{a}$). Consider the
operator $D_a = D$ with domain given by
  $$  \text{dom} (D_a) = \{ \varphi \in H^1 (X, S) \bigm| \Pi_{-a} (\varphi
|_{\bX} ) =
     0 \}. $$

        \proclaim{\protag{A.1} {Lemma}}
 $D_a$ is a closed symmetric extension of $D$, and its adjoint $D_a^*$ is given
by $D$ with domain
  $$ \text{dom} (D_a^*) = \{ \varphi \in H^1 (X, S) \bigm| \Pi_a (\varphi
     |_{\bX} ) = 0 \}. $$
        \endproclaim

        \demo{Proof}
 Proceeding in the same way as in [APS1], we can construct a two-sided
parametrix
  $$  R:\ C^\infty(X, S) \rightarrow C^\infty(X, S; \Pi_{-a}) $$
such that $DR - \operatorname{Id}$ and $RD - \operatorname{Id}$ are smoothing
operators and
  $$  R:\ H^l(X, S) \rightarrow H^{l+1} (X, S) \ \ (l \geq 0). $$
Thus if $\varphi_n \in \text{dom} (D_a)$ such that $\varphi_n \rightarrow
\varphi, \ D\varphi_n \rightarrow \psi$ in $L^2$, the existence of the
paramatrix $R$ shows that in fact $\varphi \in H^1(X, S)$ and $\varphi_n
\rightarrow \varphi$ in $H^1(X, S)$. By the continuity of the restriction map
  $$  r: \ H^1(X, S) \rightarrow H^{1/2}({\bX}, S|_{\bX}) \rightarrow
L^2({\bX}, S|_{\bX}), $$
$\varphi \in \text{dom} (D_a)$ and $D_a \varphi = \psi$. This shows that
$D_a$ is closed.

To show $D_a$ is symmetric, it suffices to prove the statement about $D_a^*$.
Integration by parts gives, for all $\varphi, \psi \in C^\infty(X, S)$,
  $$ (D\varphi, \psi ) - (\varphi, D\psi) = \int_{\bX} \langle J(\varphi
     |_{\bX}), \psi |_{\bX} \rangle
     \;\vbox{\hbox{\text{def}}\vskip-7pt\hbox{\ =}} \;\bigl(J(\varphi
     |_{\bX}), \psi |_{\bX}\bigr)_{\bX}.  \tag{A.2} $$
Again, the continuity of the restriction map $r$ shows that \thetag{A.2}
actually holds for all $\varphi, \psi \in H^1(X, S)$.

Let $D_{-a}$ denote $D$ with domain
  $$ \text{dom} (D_{-a}) = \{ \varphi \in H^1 (X, S) \bigm| \Pi_a (\varphi
|_{\bX}) =
     0 \}. $$
Then, for all $\varphi \in \text{dom} (D_a),\ \psi \in \text{dom} (D_{-a})$,
  $$ \align J(\varphi |_{\bX}) & =  J(\text{Id} - \Pi_{-a}) (\varphi |_{\bX})
\\ & =
      \Pi_a J (\varphi |_{\bX}) \\ \psi |_{\bX} & =  (\text{Id} - \Pi_{a})
(\psi
     |_{\bX}) \endalign $$
Thus $(J(\varphi |_{\bX}), \psi |_{\bX})_{\bX} = 0$ and \thetag{A.2} shows that
$D_{-a} \subset D_a^*$.

The equality $D_a^* = D_{-a}$ requires considerably more effort. Let
  $$ L^2_{int}(X,S) = \{ \varphi \in L^2(X, S) \bigm|\text{dist(supp}
     \,\varphi, \partial X) \geq \frac{1}{3} \}, $$
and
  $$  L^2_{bd}(X,S) = \{ \varphi \in L^2(X, S) \bigm|\text{supp}\, \varphi
\subset
     [0, \frac{2}{3}] \times {\bX} \}. $$
Then $L^2 (X, S) = L^2_{int}(X,S) + L^2_{bd}(X,S)$ and we just have to specify
$D_a^*$ restricted to each of the subspaces.

Clearly for $\psi \in L^2_{int}(X,S) \cap \text{dom} (D_a^*)$, we have $D_a^*
\psi = D\psi$ and
  $$  L^2_{int}(X,S) \cap \text{dom} (D_a^*) = L^2_{int}(X,S) \cap H^1(X,S). $$

The subspace $L^2_{bd}(X,S)$ splits further:
  $$  L^2_{bd}(X,S) = L^2([0, \frac{2}{3}], K_{\partial X}(a)) \oplus L^2([0,
     \frac{2}{3}], H_{\partial X} (a)), $$
where $K_{\partial X}(a), \ H_{\partial X} (a)$ are defined in (1.1).
Moreover, $D_a$ is diagonal with respect to this splitting. Now restricted to
$L^2([0, \frac{2}{3}], K_{\partial X}(a)), \ D_a = J(\partial_u + A)$, with
$A$ a symmetric endomorphism of $K_{\partial X} (a)$ which anticommutes with
$J$, and the boundary condition at $u = 0$ is $\varphi |_{u=0} = 0$. Clearly
then, $D_a^* = D_{-a}$ on $L^2([0, \frac{2}{3}], K_{\partial X}(a))$.

On the other hand, for $D_a$ restricted $L^2([0, \frac{2}{3}], H_{\partial X}
(a))$, the construction in [APS1] actually gives bounded inverse $R_a$ for
$D_a$ and $R_{-a}$ for $D_{-a}$. From
  $$  (D_a \varphi, \psi ) = (\varphi, D_{-a} \psi) $$
for $\varphi \in \text{dom} (D_a), \ \psi \in \text{dom} (D_{-a})$, we obtain,
by continuity, $R_a^* = R_{-a}$. Since adjoints commute with inverses, the
lemma is established, for the discussion above shows that $D_a^* \subset
D_{-a}$.
        \enddemo

{}From the lemma it is clear that $D_a$ is in general not self-adjoint so we
need to consider self-adjoint extensions of $D_a$. Suppose $D_s$ is such a
self-adjoint extension, then $D_a \subset D_s \subset D_a^*$, i.e. $D_s = D$
with
  $$  \text{dom} (D_a) \subset \text{dom} (D_s) \subset \text{dom} (D_a^*).
     \tag{A.3} $$

Recall our notation from \S 1. We have $K_{\partial X}(a) = \text{Im} (\Pi_{-a}
- \Pi_a)$ splits into the $(\pm i)$-eigenspace of $J$ (Cf (1.1)):
  $$  K_{\partial X}(a) = K_{\partial X}^+(a) \oplus K_{\partial X}^-(a). $$

        \proclaim{\protag{A.4} {Lemma}}
 We have $\dim K_{\partial X}^+(a) = \dim K_{\partial
X}^-(a)$.
        \endproclaim

        \demo{Proof}
  This is a consequence of the cobordism invariance of index.  Alternatively,
it follows from the Atiyah-Patodi-Singer index formula, as follows. First of
all, by the symmetry of $\text{spec}D_{\partial X}$, we just need to show
that for $a$ less than the smallest nonzero eigenvalue of $D^2_{\partial X}$.
Namely, $\dim K_{\partial X}^+ = \dim K_{\partial X}^-$, where $K_{\partial
X}^{\pm}$ are the $\pm i$-eigenspace of $J$ restricted to $\ker D_{\partial
X}$. Applying the APS index formula to $D_a$ yields
  $$ \dim L=\frac{\dim \ker D_{\partial X}}{2}, \tag{A.5} $$
where $L\subset \ker D_{\partial X}$ is the subspace of limiting values of
the extended $L^2$-solutions of $D$ (see [APS1]). Alternatively, $L=\Pi
r(\ker D_a^*)=\Pi r(\ker D_{-a})$, where $\Pi$ is the orthogonal projection
onto $\ker D_{\partial X}$. From \thetag{A.2}, together with \thetag{A.5},
we see that $L$ is a ``lagrangian" subspace of $(\ker D_{\partial X},\
(\cdot,\ \cdot)_{\bX},\ J)$: $(J\alpha, \ \beta)_{\bX}=0$ for all $\alpha,
\beta \in L$. This shows that the $(+i)$-eigenspace of $J$ has the same
dimension as the $(-i)$-eigenspace.
        \enddemo

We now denote $h^+(a)=\dim  K_{\partial X}^+(a)$.

        \proclaim{\protag{A.6} {Proposition}}
 There is a one-one correspondence
  $$  \{ \text{self-adjoint extensions of}\ D_a \} \longleftrightarrow \{
     \text{unitary maps}\ T:\ K_{\partial X}^+(a)\rightarrow K_{\partial
     X}^-(a) \}. $$
For a unitary map $T$, its corresponding self-adjoint extension $D(a,T)$ is
given by $D$ with
  $$ \text{dom} (D(a, T)) = \{ \varphi \in H^1 (X, S) \bigm| (\Pi_a +
     \Pi_T)(\varphi |_{\bX}) = 0 \}, $$
where $\Pi_T$ is the orthogonal projection onto the graph of $T$ in
$K_{\partial X}(a)$.
	\endproclaim

        \demo{Proof}
 Any self-adjoint extension of $D_a$ is given by $D_s = D$ with domain
satisfying \thetag{A.3}. Thus
  $$  r(\text{dom} (D_a)) \subset r(\text{dom} (D_s)) \subset r(\text{dom}
     (D_a^*)). $$
Or
  $$ r(\text{dom} (D_a)) \subset r(\text{dom} (D_s)) \subset r(\text{dom}
     (D_a)) \oplus K_{\partial X}(a). $$

{}From \thetag{A.2},
  $$  (J(\varphi |_{\bX}), \psi |_{\bX})_{\bX} \equiv 0 \tag{A.7} $$
for all $\varphi, \psi \in$ dom$(D_s)$, or equivalently, for all $\varphi
|_{\bX}, \psi |_{\bX} \in r(\text{dom} (D_s))$. Since $D_a$ is symmetric,
\thetag{A.7} is automatically satisfied on $r(\text{dom} (D_s))$. Let $L =
r(\text{dom} (D_s)) \cap K_{\partial X}(a)$ be a subspace of $K_{\partial
X}(a)$. Then \thetag{A.7} shows that $L$ is an ``isotropic" subspace of
$(K_{\partial X}(a), J)$. Since $D_s$ is self-adjoint, $L$ must be maximal
isotropic, hence ``lagrangian". Now it is a little linear algebra to show
that there is a one-one correspondence
  $$ \{ \text{lagrangian subspace}\ L\ \text{of}\ (K_{\partial X}(a), J)
     \leftrightarrow \{ \text{unitary map}\ T:\ K_{\partial
     X}^+(a)\rightarrow K_{\partial X}^-(a) \} $$
given by $L =$ the graph of $T$. This shows one way of the correspondence.
But the other direction is completely similar to the proof of \theprotag{A.1}
{Lemma}.
        \enddemo

        \rem
 This is very similar to von Neumann's theory of deficiency indexes which
completely characterizes self-adjoint extensions of a closed symmetric
operator.
        \endrem

        \rem
 Formally, for $D$ with domain $C_0^\infty(X, S)$, there is also a one-one
correspondence
  $$ \align \{ \text{self-adjoint extensions of}\ D\} & \leftrightarrow  \{
     \text{unitary maps}:\ H^+_{\partial X} \rightarrow H^-_{\partial X} \}
     \\ & \leftrightarrow  \{ \text{lagrangian subspaces of}\ H_{\partial X}
     = L^2({\bX}, S|_{\bX}) \}. \endalign $$
However, one loses the ellipticity in this generality.
        \endrem

Thus, given $a \not\in \text{spec} D_{\partial X}^2$ positive and $T:\
K_{\partial X}^+(a) \rightarrow K_{\partial X}^-(a)$ an isometry (unitary),
the operator~$D(a,T)$ is self-adjoint, and, as we mentioned earlier, elliptic
in a generalized sense. We will not, however, go into the discussion of the
ellipticity of $D(a,T)$, but instead, derive some of its consequences from
the study of the heat kernel, $e^{-tD^2(a, T)}$.

For this purpose, we first consider the situation on the infinite half cylinder
$R_+ \times {\bX}$. In this case, $D = J(\partial_u + D_{\partial X})$ and we
have a global decomposition.
  $$ \align L^2(R_+ \times {\bX},S) & =  L^2(R_+, L^2({\bX}, S|_{\bX}))  \\ & =
      L^2(R_+, K_{\partial X} (a)) \oplus L^2(R_+, H_{\partial X} (a)).
     \endalign $$
Since both $D$ and the boundary condition are diagonal with respect to this
decomposition, \break $e^{-tD^2(a,T)} = E_{<a}(t) + E_{>a}(t)$ splits into
two pieces as well. As the boundary condition on $L^2(R_+, H_{\partial X}
(a))$ is completely analogous to the APS boundary condition, $E_{>a}(t)$ can
be given an explicit formula. Let $\{ \varphi_\lambda; \lambda \in$
spec$D_{\partial X},\ \lambda > \sqrt{a} \}$ be an orthonormal basis for
Im$\Pi_a$ consisting of eigensections of $D_{\partial X}$. Then the same
construction in~\cite{APS} gives
  $$ \multline E_{>a}(t) = \sum_{\lambda > \sqrt{a}}
     \frac{e^{-\lambda^2t}}{\sqrt{4 \pi
     t}}(e^{-(u-v)^2/4t}-e^{-(u+v)^2/4t})\,\varphi_{\lambda}\otimes
     \varphi_{\lambda}^*  \\
      + \Bigl\{\frac{e^{-\lambda^2t}}{\sqrt{4 \pi
     t}}(e^{-(u-v)^2/4t}+e^{-(u+v)^2/4t})
       -\lambda e^{\lambda(u+v)}\text{erfc}(\frac{u+v}{2\sqrt{t}}+ \lambda
     \sqrt{t})\Bigr\}\,J\varphi_{\lambda}\otimes J\varphi_{\lambda}^* .
     \endmultline $$
On the other hand, there is no explicit formula for $E_{<a}(t)$. But it is
reduced to a heat kernel on the half line $R_+$, with $L^2$ boundary condition
at $\infty$ and local elliptic condition at $0$:
  $$ \aligned (\partial_t - \partial_u^2+A^2) E_{<a}(t,u,v) & = 0 \\
     E_{<a}|_{t=0} & = \text{Id} \\ \Pi_TE_{<a}|_{u=0} & = 0\\ \ J\Pi_T
     J(\partial_u + A)E_{<a}|_{u=0} & = 0, \endaligned $$
with $A = D_{\partial X}|_{K_{\partial X}(a)}$ a finite dimensional symmetric
endomorphism.

To discuss the heat kernel on $X$, we use the patching construction of
[APS1]. More precisely, let $\rho (a,b)$ be an increasing $C^\infty$ function
on $R$ such that $\rho = 0$ for $u \leq a$ and $\rho = 1$ for $u \geq b$.
Define
  $$ \align \phi_1 = & \rho(\frac{1}{6}, \frac{2}{6}), \ \ \psi_1 =
     \rho(\frac{3}{6}, \frac{4}{6}) \\ \phi_2 = & 1- \rho(\frac{5}{6}, 1), \
     \psi_2 = 1-\psi_1. \endalign$$
These extend to smooth functions on $X$ in an obvious way. Let $\tilde{D}$ be
the Dirac operator on the double of $X$. Then
  $$ e = \phi_1 e^{-t \tilde{D}^2}\psi_1 + \phi_2 (E_{<a}(t) + E_{>a}(t))
     \psi_2 $$
is a parametrix for the heat operator $\partial_t + D^2(a,T)$, and
  $$  e^{-tD^2(a,T)} = e + \sum_{m=1}^\infty (-1)^m c_m * e, \tag{A.8} $$
where $*$ denotes the convolution of kernels, $c_1 = (\partial_t +
D^2(a,T))e$, and $c_m = c_{m-1} * c_1,\ m \geq 2$. It follows that for $t >
0, \ e^{-tD^2(a,T)}$ is a $C^\infty$ kernel which differs from $e$ by an
exponentially small term as $t \rightarrow 0$.

        \proclaim{\protag{A.9} {Lemma}}
 \rom(i\rom) Both $e^{-tD^2(a,T)}$ and $D(a,T)e^{-tD^2(a,T)}$ are trace class
for $t>0$.

\rom(ii\rom) As $t \rightarrow 0$,
  $$  \text{Tr} (e^{-tD^2(a,T)}) \sim \sum_{j=0}^\infty a_j (D(a,T))
     t^{(j-n)/2}, $$
and
  $$  \text{Tr} (D(a,T)e^{-tD^2(a,T)}) \sim \sum_{j=0}^\infty b_j (D(a,T))
     t^{(j-n-1)/2}, $$
with $a_j,\ b_j$ given by integral of local densities computable from the
(total) symbol of $D$ and boundary conditions.
        \endproclaim

        \demo{Proof}
 (i) Since for $t >0, \ e^{-tD^2(a,T)}$ is smooth, it is Hilbert-Schmidt. Now
the semi-group properties show that $e^{-tD^2(a,T)} = e^{-\frac{t}{2}D^2(a,T)}
\circ e^{-\frac{t}{2}D^2(a,T)}$ is a product of Hilbert-Schmidt operators,
hence trace class. Similarly for $D(a,T) e^{-tD^2(a,T)}$.

(ii) From (i) and Lidskii's theorem
  $$  \text{Tr} (e^{-tD^2(a,T)}) = \int_X \text{tr} (e^{-tD^2(a,T)}) (x,x)
     dx. $$
For the asymptotic expansion we may replace $e^{-tD^2(a,T)}$ by its parametrix
$e$. The asymptotic expansion for $e$ follows from its explicit construction,
as in [APS1].
        \enddemo

        \proclaim{\protag{A.10} {Corollary}}
 The spectrum of $D(a,T)$ consists of eigenvalues of finite multiplicities
satisfying Weyl's asymptotic law:
  $$ \align N(\lambda) & =  \# \{ \lambda_j \, \bigm| \, |\lambda_j| \leq
\lambda
     \} \\ & =  \frac{\text{vol} (X)}{(4\pi)^{n/2} \Gamma(\frac{n}{2} + 1)}
     \lambda^n + o(\lambda^n) \qquad \text{as} \ \lambda \rightarrow \infty.
     \endalign $$
        \endproclaim

Thus, the eta function
  $$ \eta(s, D(a,T)) = \sum_{\lambda_j \not= 0} \text{sign} \lambda_j
     |\lambda_j|^{-s} $$
is well-defined for Re$\,s > n$. Further by Mellin transform,
  $$  \eta(s, D(a,T)) = \frac{1}{\Gamma (\frac{s+1}{2})} \int_0^\infty
     t^{(s-1)/2} \text{Tr} (D(a,T)e^{-tD^2(a,T)}) dt. \tag{A.11} $$
And Lemma~\thetag{A.9} shows that $\eta(s, D(a,T))$ admits a meromorphic
continuation to the complex plane with only simple poles.

        \proclaim{\protag{A.12} {Proposition}}
 $\eta(s, D(a,T))$ is actually holomorphic in Re$\,s >
-\frac{1}{2}$. Therefore the eta invariant $\eta(a,T) = \eta(0, D(a,T))$ is
well-defined. Moreover
  $$  \eta(a,T) = \frac{1}{\sqrt{\pi}} \int_0^\infty t^{-1/2} \text{Tr}
     (D(a,T)e^{-tD^2(a,T)}) dt. $$
        \endproclaim

        \demo{Proof}
 It suffices to show that
  $$ \text{Tr} (D(a,T)e^{-tD^2(a,T)}) = O(1) \ \text{as} \ t \rightarrow 0. $$
The same argument as in the proof of Lemma~\thetag{A.9} shows that
  $$ \split \text{Tr} (D(a,T)e^{-tD^2(a,T)}) & = \int_X \text{tr} (D_x
     e(t,x,x')|_{x=x'}) dx + O(e^{-c/t})  \\ & = \int_X \text{tr}
     (D_x e^{-t\tilde{D}^2}(x,x')|_{x=x'})\psi_1(x) dx  \\ &\qquad \qquad   +
     \int_{R_+ \times {\bX}} \text{tr} (D_x (E_{<a}(t) + E_{>a}(t))|_{x=x'}
     \psi_2(x) dx  + O(e^{-c/t}). \endsplit \tag{A.13} $$

The local cancellation result for closed manifold gives
  $$  \text{tr} (D_x e^{-t\tilde{D}^2}(x,x')|_{x=x'}) = O(t^{1/2}) $$
uniformly in $x$. Therefore the first term in \thetag{A.13} is $O(t^{1/2})$.

For the second term, a straightforward calculation shows that
  $$ \int_{\bX} \text{tr} (D_x E_{>a}(t)|_{x=x'}) \equiv 0. $$
Also tr$(JAE_{<a}(t)) \equiv 0$ since $JA = -AJ$. Thus
  $$ \text{Tr} (D(a,T)e^{-tD^2(a,T)}) = \int_{R_+ \times {\bX}} \text{tr}
     [J\partial_u E_{<a}(t)|_{u=v}] \psi_2(u) du dy + O(t^{1/2}). $$

Since $E_{<a}(t)$ is the heat kernel of an elliptic local boundary value
problem on $R_+$, we have
  $$ E_{<a}(t,u,v) = \frac{e^{-(u-v)^2/4t}}{\sqrt{4\pi t}} (1 +
     b_1(T,u,v)t^{1/2} + O(t)) $$
uniformly in $u,\ v$. Therefore
  $$ J\partial_u E_{<a}(t)|_{u=v} = \frac{1}{\sqrt{4\pi }}J\partial_u
     b_1(T,u,v)|_{u=v} + O(t^{1/2}), $$
and our claim follows.
        \enddemo

We now turn to the variation of eta invariants. For our purpose we are going
to work in complete generality. So let $P(z)$ be a family of operators
satisfying:\medskip

\property{Ha}{$P(z)$~is a smooth family of (unbounded) self-adjoint operators
on $L^2(X,S)$ with dom$\, (P(z))$ independent of the parameter $z$;}
\smallskip
\property{Hb}{The heat semi-group $e^{-tP^2(z)}\ (t>0)$ is a smooth family of
smoothing operators, i.e. the heat kernel is given by smooth functions on $X$
depending smoothing on $z$.}

        \proclaim{\protag{A.14} {Lemma}}
 For a family satisfying \rom(Ha\rom), \rom(Hb\rom), we have
  $$  \frac{\partial}{\partial z} \text{Tr} (P(z) e^{-tP^2(z)}) = (1+2t
     \frac{\partial}{\partial t}) \text{Tr} (\dot{P}(z) e^{-tP^2(z)}). $$
        \endproclaim

        \demo{Proof}
 First of all,
  $$ \frac{\partial}{\partial z} \text{Tr} (P(z) e^{-tP^2(z)}) = \text{Tr}
     (\dot{P}(z) e^{-tP^2(z)}) + \text{Tr} (P(z) \frac{\partial}{\partial z}
     e^{-tP^2(z)}). $$
To compute $ \frac{\partial}{\partial z} e^{-tP^2(z)}$, we apply the heat
operator:
  $$ (\frac{\partial}{\partial t} + P^2(z)) \frac{\partial}{\partial z}
     e^{-tP^2(z)}) = [P^2(z), \frac{\partial}{\partial z}] e^{-tP^2(z)}). $$
Now, with the initial condition of the heat equation and dom$(P(z))$
independent of $z$, Duhamel's principle gives
  $$ \frac{\partial}{\partial z} e^{-tP^2(z)}) = \int_0^t e^{-(t-s)P^2(z)}
     [P^2(z), \frac{\partial}{\partial z}] e^{-sP^2(z)} ds. $$
Consequently
  $$ \align \text{Tr} (P(z) \frac{\partial}{\partial z} e^{-tP^2(z)}) & =
     -2t\text{Tr} (\dot{P}(z) P^2(z) e^{-tP^2(z)}) \\ & =  2t
     \frac{\partial}{\partial t} \text{Tr} (\dot{P}(z) e^{-tP^2(z)}).
     \endalign $$
This finishes the proof.
     \enddemo

We now consider the variation of eta function $\eta (s, P(z))$ defined by
\thetag{A.11}. For it to be well-defined we make the following additional
assumption: \medskip
\property{Hc}{There is a uniform asymptotic expansion of Tr$(P(z)
e^{-tP^2(z)})$ at $t=0$:
  $$  \text{Tr}(P(z) e^{-tP^2(z)}) \sim \sum_{j \geq -N} a_j (P(z)) t^{j/d}, $$
and $a_j(P(z))$ are smooth in $z$.}

        \proclaim{\protag{A.15} {Lemma}}
 Let $P(z)$ be a family of operators satisfying \rom(Ha\rom), \rom(Hb\rom),
and \rom(Hc\rom).  Furthermore, assume that $\dim \ker P(z)$ is constant.
Then for Re$\,s > N$, we have
  $$  \frac{\partial}{\partial z} \eta(s,P(z)) = -\frac{s}{\Gamma
     (\frac{s+1}{2})} \int_0^\infty t^{(s-1)/2} \text{Tr} (\dot{P}(z)
     e^{-tP^2(z)}) dt. $$
        \endproclaim

        \demo{Proof}
 By (Hb), $P(z)$ all have discrete spectrum. It follows from the assumption on
$\dim \ker P(z)$ that Tr$\, (P(z)e^{-tP^2(z)})$ is exponentially decaying,
uniformly in $z$, as $t \rightarrow \infty$. (Hc) implies that $\eta(s,P(z))$
analytically continues to a meromorphic function smooth in $z$.

Let $T > 0$ and Re$\, s > N$, By \theprotag{A.14} {Lemma},
  $$ \multline \frac{\partial}{\partial z} \int_0^T t^{(s-1)/2}
     \text{Tr}(P(z) e^{-tP^2(z)}) dt \\ = 2T^{(s+1)/2} \text{Tr} (\dot{P}(z)
     e^{-TP^2(z)}) -s \frac{\partial}{\partial z} \int_0^T t^{(s-1)/2}
     \text{Tr}(\dot{P}(z) e^{-tP^2(z)}) dt. \endmultline \tag{A.16} $$
Denote by $H(z)$ the orthogonal projection of $L^2(X,S)$ onto $\ker P(z)$.
Since $\dim \ker P(z)$ is constant, $H(z)$ depends smoothly on $z$.
Furthermore, the self-adjointness of $P(z)$ implies that
  $$ P(z) H(z) = H(z) P(z) = 0. $$
Therefore
  $$ P(z) = (\text{Id} - H(z)) P(z) (\text{Id} - H(z)), $$
and hence
  $$ \dot{P}(z) = -\dot{H} (z) P(z) (\text{Id} - H(z)) + (\text{Id} - H(z))
     \dot{P}(z) (\text{Id} - H(z)) - (\text{Id} - H(z)) P(z) \dot{H}(z). $$
Since $(\text{Id} - H(z)) e^{-tP^2(z)}$ is given by a smooth kernel decaying
exponentially in $t$ as $t \rightarrow \infty$, it follows that the right
hand side of \thetag{A.16} is absolutely convergent so we can take the limit
of \thetag{A.16} as $T \rightarrow \infty$ and exchange the limit with the
differentiation. The same discussion applies to the left hand side of
\thetag{A.16} and we obtain the lemma.
        \enddemo

An immediate consequence of the lemma is that when $\eta(s,P(z))$ are all
regular at $s=0$,
  $$ \frac{\partial}{\partial z} \eta(P(z)) = -\frac{2}{\sqrt{\pi}}
     \text{LIM}_{t \rightarrow 0} t^{1/2} \text{Tr} (\dot{P}(z) e^{-t
     P^2(z)}), $$
where $\text{LIM}_{t \rightarrow 0}$ means taking the constant term in the
asymptotic expansion at $t = 0$.

Now define
  $$ \xi (P(z)) = \frac{\eta (P(z)) + \dim \ker P(z)}{2}. $$

        \proclaim{\protag{A.17} {Proposition}}
 Let \rom(Ha\rom), \rom(Hb\rom), \rom(Hc\rom) hold for $P(z)$. Then $\xi
(P(z))\pmod1$ defines a smooth function and
  $$  \frac{d}{dz} \xi (P(z)) = -\frac{1}{\sqrt{\pi}} \text{LIM}_{t
     \rightarrow 0} t^{1/2} \text{Tr} (\dot{P}(z) e^{-t P^2(z)}). $$
        \endproclaim

        \demo{Proof}
 Choose a $c > 0$ such that $c$ is not in the spectrum of $P(z)$ for all $z$ in
a small neighborhood. Let $\Pi_c(z)$ be the orthogonal projection onto the
space spanned by all eigensections with eigenvalues $\lambda$ satisfying
$|\lambda | < c$. Define a new family
  $$ P^c(z) = P(z) (\text{Id} - \Pi_c (z)) + \Pi_c (z). $$
Namely one replaces by $1$ all eigenvalues $\lambda$ of $P_B(z)$ satisfying
$|\lambda | < c$ and leave the rest unchanged. Therefore $P^c(z)$ is clearly
invertible, $\xi (P^c(z)) = \frac{1}{2} \eta (P^c(z))$ is smooth, and
  $$  \frac{d}{dz} \xi (P^c(z)) = -\frac{1}{\sqrt{\pi}} \text{LIM}_{t
     \rightarrow 0} t^{1/2} \text{Tr} (\dot{P}^c(z) e^{-t (P^c(z))^2}). $$
Now
  $$ \xi (P(z)) = \xi (P^c(z)) + \frac{\sum_{\lambda \in spec P(z), |\lambda|
     < c} (\text{sign} \lambda - 1)}{2}, $$
here
  $$ \text{sign} \lambda = \cases \hphantom{-}1 ,&\text{if $\lambda \ge0$}
     ;\\-1,&\text{if $\lambda <0$} .\endcases $$
Clearly then
  $$ \xi (P_B(z)) \equiv \xi (P_B^c (z)) \ \ \text{mod} \ \ {\Bbb Z}. $$
On the other hand,
  $$ e^{-t(P^c(z))^2} = e^{-t P^2(z)} + \text{finite rank}, $$
and
  $$ \dot{P}^c (z) = \dot{P} (z) + \text{finite rank}, $$
which implies that
  $$ \text{Tr} (\dot{P}^c (z) e^{-t(P^c(z))^2}) = \text{Tr} (P(z) e^{-t
     P^2(z)}) + O(1). $$
Therefore
  $$ \text{LIM}_{t \rightarrow 0} t^{1/2} \text{Tr} (\dot{P}^c(z) e^{-t
     (P^c(z))^2}) = \text{LIM}_{t \rightarrow 0} t^{1/2} \text{Tr}
     (\dot{P}(z) e^{-t P^2(z)}). $$
        \enddemo

Finally, we point out that although the $L^2$-norm on $L^2(X,S)$ depends on
the metric, a smooth family of metrics gives rise to a smooth family of
equivalent norms. Therefore the resulting trace functional on $L^2(X,S)$ is
independent of the metric change.

\enddocument
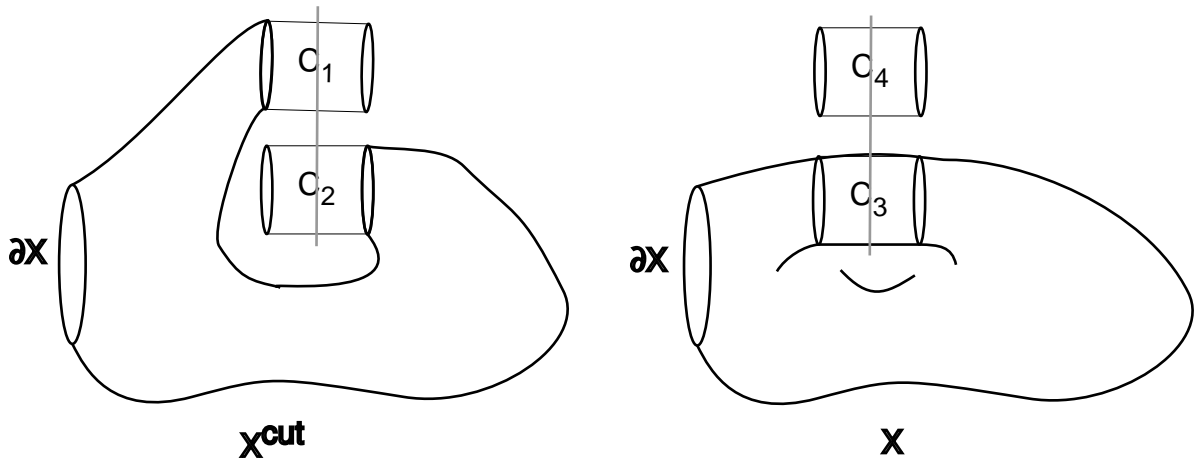
